\documentclass{aa}

\usepackage{newtxtext,newtxmath}
\usepackage[T1]{fontenc}
\usepackage[dvipsnames]{xcolor}
\usepackage[normalem]{ulem}
\usepackage[bottom]{footmisc}
\usepackage{lineno}
\usepackage{amsfonts, dsfont}
\usepackage{savesym}
\usepackage[nointegrals]{wasysym}
\usepackage[normalem]{ulem}
\usepackage[hang]{subfigure}
\usepackage{hyperref}
\usepackage{accents}
\usepackage{threeparttable, booktabs}
\savesymbol{tablenum}
\usepackage{physics}
\usepackage{siunitx}
\usepackage{mathtools}
\restoresymbol{SIX}{tablenum}
\usepackage{changes}
\usepackage{float}

\DeclareSIUnit{\erg}{erg}
\DeclareSIUnit{\jansky}{Jy}
\DeclareSIUnit{\parsec}{pc}
\DeclareSIUnit{\yr}{yr}
\DeclareSIUnit{\hr}{hr}
\DeclareSIUnit\msun{M\textsubscript{\astrosun}}
\DeclareSIUnit{\radian}{rad}
\DeclareSIUnit{\pixel}{px}
\DeclareSIUnit{\day}{d}
\DeclareSIUnit{\arcsectxt}{as}
\DeclareSIUnit{\arcsec}{arcsec}
\DeclareSIUnit{\sr}{sr}
\DeclareSIUnit{\correlation}{correlation}
\DeclareSIUnit{\strain}{strain}

\newcommand{\vct}[1]{\accentset{\rightharpoonup}{#1}}
\newcommand{\mtx}[1]{\mathbf{#1}}

\newcommand{\gw}{\mathrm{gw}}

\newcommand{\GWB}{\mathrm{GWB}}
\newcommand{\lmax}{\ell_\mathrm{max}}
\newcommand{\sreg}{s_\mathrm{reg}}

\newcommand{\mpifr}{Max-Planck-Institut f{\"u}r Radioastronomie, Auf dem H{\"u}gel 69, 53121 Bonn}
\newcommand{\SPA}{School of Physics and Astronomy, Monash University, Clayton VIC 3800, Australia}
\newcommand{\OzGravMonash}{OzGrav: The ARC Centre of Excellence for Gravitational Wave Discovery, Clayton VIC 3800, Australia}
\newcommand{\ozgrav}{OzGrav: The Australia Research Council Centre of Excellence for Gravitational Wave Discovery}
\newcommand{\jbca}{Jodrell Bank Centre for Astrophysics, University of Manchester,\\ Department of Physics and Astronomy, Alan-Turing Building, Oxford Street, Manchester M13 9PL, UK}
\newcommand{\vblt}{Department of Physics and Astronomy, Vanderbilt University, 2301 Vanderbilt Place, Nashville, TN 37235, USA}

\begin{document}

\title{Optimising gravitational-wave sky maps for pulsar timing arrays}

\author{Kathrin Grunthal \inst{1}\thanks{kgrunthal@mpifr-bonn.mpg.de} \and 
        David J.~Champion \inst{1} \and
        Eric Thrane \inst{2,3} \and
        Rowina S.~Nathan \inst{2,3} \and
        Michael Kramer\inst{1,4} \and
        Matthew T.~Miles\inst{5,6}
        }

\institute{
    $^{1}$\mpifr\\
    $^{2}$\SPA\\
    $^{3}$\OzGravMonash\\
    $^{4}$\jbca\\
    $^{5}$\vblt\\
    $^{6}$\ozgrav
}

\date{\today}

\abstract
    {Pulsar timing arrays (PTAs) have recently reported compelling evidence for the presence of a gravitational-wave background signal. Mapping the gravitational-wave background is key to understanding how it is formed, since anisotropy is a tracer for, for example, a supermassive black hole binary origin.}
    {In this work we refine the frequentist regularised gravitational-wave mapping analysis developed in our previous work (as part of the MeerKAT PTA 4.5-year data release). We derive a point-spread function describing the angular resolution of a PTA.}
    { We investigate how the point spread function changes for different PTA constellations and determine the best possible angular resolution achievable within our framework. Using simulated data, we demonstrate that previous methods do not capture the actual resolution -- especially in regions of the sky with a high density of pulsars.}
    {We propose an improved scheme that accounts for a variable local resolution and test it using realistic simulations of the latest MeerKAT dataset. We demonstrate that we are able to identify a continuous gravitational wave signal in a region with good pulsar sky coverage with approximately a factor of two increase in significance compared to our previous method. }
    {}


\maketitle

\section{Introduction} \label{sec:intro}

The gravitational wave (GW) spectrum accessible with current and future observatories is commonly divided into three different regimes reflecting the observing band of these detectors. In contrast to the individually resolvable GWs in the intermediate- and high-frequency regime ($> \SI{10}{\micro\hertz}$), the primarily targeted GW signal in the nanohertz regime is the gravitational wave background (GWB) \citep{Rosado_2015}.

It is likely that this GWB is caused by the incoherent superposition of GWs emitted by the population of supermassive black hole binaries \citep{Rajagopal_1995,Allen_1997}. Supermassive black holes are thought to be found in centres of galaxies \citep{Kormendy_1995}, and so a supermassive black hole binary is likely to form during a galaxy merger \citep{Begelman_1980}.
Historically, the inspiral of circular supermassive black hole binaries was expected to stall \citep[`final parsec problem'][]{Milosavljevic_2003}, but various theories have been put forward that allow the supermassive black holes to get so close that they become GW-driven \citep{HolleyBockelmann_2015,Gualandris_2017,Ryu_2018}.
Such a supermassive black hole binary-originating GWB would create an anisotropic distribution of GW power across the sky, once due to the finite number of supermassive black hole binaries contributing to the signal \citep{Allen_1997,Mingarelli_2013,Gair_2014,Mingarelli_2017}, but also as it is expected that the supermassive black hole binary population follows the large-scale cosmic structure \citep{Burke-Spolaor_2019}. Additionally, individual close-by 'loud' supermassive black hole binaries are expected to stand out from the quasi-isotropic background \cite{Sesana_2004,Yardley_2010,SesanaVecchio_2010}, so-called hotspots. 

A nanohertz GWB signal can also be caused by early-Universe processes \citep{Maggiore_2000}, such as phase transitions \citep{Caprini_2020, Hindmarsh_2021}, cosmic string collapse \citep{Hindmarsh_1995, Saikawa_2017}, or primordial GWs from quantum fluctuations that were redshifted by inflation \citep{Guzzetti_2016, Lasky_2016, Yuan_2021, Domenech_2021}. For most of these models, the GWB is expected to be isotropic.

This makes the nanohertz GW regime interesting -- either offering a unique window onto galaxy evolution studies and/or constraining processes in the early Universe. For nanohertz frequencies, a suitable GW detector needs to be galaxy-sized. This is realised in pulsar timing arrays (PTAs) \citep{Sazhin_1978,Detweiler_1979}. A PTA consists of a group of millisecond pulsars that are regularly monitored over years to decades, usually using radio telescopes. 
Gravitational wave detection with PTAs is based on the exceptional rotational stability of these pulsars. Due to the clock-like predictability of the pulse times-of-arrival (ToAs), even small-scale deviations caused by GWs can be detected. Pulsar timing residuals are obtained by subtracting the ToA prediction of the best-fitting timing model from the recorded ToAs. 
Residuals caused by GWs are then detected by searching for quadrupolar correlations across the residual series of all pulsars in the PTA. 
The stochastic GWB is characterised as a noisy red process \citep{AllenRomano_1999} that exhibits angular correlation according to the Hellings-Downs curve \citep{HellingsDowns}.

In recent years, different PTA collaborations have independently reported compelling evidence for the presence of such a GWB signal in their datasets \citep{EPTA_DR2_GWB,NANOGrav15yObs,PPTA_GWB,MPTA2025_GWB,CPTA_GWB}. Naturally, one of the next key steps is determining the origin of this signal. A key method of discriminating the different scenarios described above is calculating GW sky maps and determining the level of anisotropy across the sky map. A significant anisotropy is an indicator of an astrophysical GWB. Simultaneously, a hotspot is a smoking gun for the presence of an individually resolvable supermassive black hole binary, a signal that is yet to be detected in PTA datasets. Hence, the search for anisotropy is of increased interest to the PTA community and has gained significant momentum.

Based on the sky mapping methods built for ground- and space-based GW detectors \citep{Mitra_2008,Thrane_2009,Banagiri_2021}, several approaches to obtain GW power sky maps from PTA data were developed. Those methods, both frequentist \citep{Thrane_2009,Pol_2022,MPTA2025_aniso} and Bayesian \citep{Mingarelli_2013,Taylor_2013,Cornish2014,Taylor_2020}, analyse and quantify the level of anisotropy by expanding the GW power in some choice of sky-resolving basis, and constraining the basis coefficients. The spherical harmonic expansion is used most widely, either linearly, or in form of the square-root spherical harmonics to ensure positivity of the recovered sky map. The application of these methods to real PTA datasets was performed in \cite{Taylor_2015,Agazie_2023}, and \cite{MPTA2025_aniso}, our recent anisotropy study as part of the MeerKAT PTA (MPTA) 4.5-year data release \citep{MPTA2025_data+noise}. In that work we have presented a framework to characterise the anisotropy of the GWB that improved the spherical harmonic frequentist approach \citep{Thrane_2009,Pol_2022} by installing a regularised maximum likelihood estimation that includes GWB self-noise. 

A key point of calculating GW sky maps is to account for the diffraction limit of the PTA, caused by the limited number of pulsars used to construct the map. All recent works on PTA GW power map making \citep{Romano_2017,Agazie_2023,MPTA2025_aniso} based on spherical harmonic expansions handle the diffraction limit of a PTA by imposing that the maximum number of spherical harmonics modes used in the expansion should not surpass the number of pulsars in the array. This approach treats the diffraction limit as a global problem, and in consequence introduces a globally defined maximum angular resolution of the resulting sky map.
However, as was shown by \cite{Boyle_Pen_2012}, the local resolving power of a PTA dataset strongly depends on the pulsar distribution across the sky. In addition, \cite{Floden_2022} demonstrated that for Earth-based GW detectors the sky localisation of single sources can be improved by including higher-order spherical harmonics beyond $\lmax$ obtained from the diffraction limit argument.
Thus, the ultimate aim of this follow-up work to \cite{MPTA2025_aniso} is to improve the currently used regularised spherical harmonics expansion scheme such that it properly reflects the local resolution of the PTA, while respecting the global analysis limit imposed by the number of pulsars. We expect that the updated scheme also provides a more precise localisation of anisotropic structures in the map, which ultimately can optimise follow-up analyses \citep{DOrazio_2023,NG15_Targets} and guide candidate vetting .

We begin by giving a brief introduction to PTA sky map analysis in Sec.~\ref{sec:methods}. The first major result of our work is then presented in Sec.~\ref{sec:sky_resolution}, in which we develop a mathematical framework to calculate the point spread function (PSF) of a PTA dataset analysed with a regularised spherical harmonics framework. We employ it to revisit the spherical harmonics expansion scheme using a hypothetical isotropic PTA, presented in Sec.~\ref{sec:rev_sph_scheme}. We first discuss two major shortcomings of the current expansion, and then present the improved expansion strategy allowing to overcome these, the second major result of this work. Finally, we test the reworked analysis set up utilising more realistic simulations of the MPTA 4.5-year dataset, and discuss the results in Sec.~\ref{sec:test_mpta}. We conclude with a summary and outlook in Sec.~\ref{sec:summary}.

\section{Methods} \label{sec:methods}

Our methods and notation closely follow our previous work \cite{MPTA2025_aniso}. Hence we only give a minimum self-consistent overview on the basics of PTA datasets and GW sky mapping from those in this work. We refer to \cite{MPTA2025_aniso} and references therein for further details, especially on the mathematical intricacies of the method.

\subsection{PTA datasets and the overlap reduction function}

A PTA dataset is a collection of ToA time series, $\{\vct{\delta t}\}$, from $N_\mathrm{psr}$ pulsars, in which a GWB signal appears as an angularly correlated stochastic process. The ToAs are characterised by a multivariate normal distribution with a covariance matrix. The total noise budget of a PTA dataset is quantified in terms of the ToA covariance matrices: 
\begin{equation}
    \mathcal{C}_{ab} = \langle\vct{\delta t}_a\vct{\delta t}_b^T\rangle .
\end{equation}
The auto-correlation part, $C_{aa}$, describes the noise processes associated with each pulsar, whereas the cross-correlation matrix,
\begin{equation}\label{eq:S2}
    S_{ab} = \vct{F}_a \; \Gamma_{ab} \mtx{\phi} \; \vct{F}_b^\dag, 
\end{equation}
depends only on the GW signal. Via the Fourier basis vectors, $\vct{F}_{a,b}$, it is usually expressed in terms of the angular correlation coefficients, $\Gamma_{ab}$ \citep{Allen_1997}, and the diagonal Fourier domain covariance matrix,
\begin{equation}
    \mtx{\phi}_{kk'} = \delta_{kk'}S(f_k)\Delta f = \frac{A_\GWB^2}{12\pi^2}\left(\frac{f}{f_\mathrm{yr}}\right)^{-\gamma_\GWB}.
\end{equation}
This matrix is built from the GWB power spectral density (PSD) $S(f_k)$, parametrised in terms of an amplitude, $A_\GWB$, and a spectral index, $\gamma_\GWB$, as well as the frequency bin width, $\Delta f = 1/T_\mathrm{obs}$, where $T_\mathrm{obs}$ denotes the total time span of the dataset, and the normalising factor $f_\mathrm{yr}= 1/\SI{1}{\yr}$. For the GWB it is expected that $\gamma_\GWB >0$, especially, if it originates from circularly inspiraling supermassive black hole binaries, $\gamma_\GWB=13/3$. A stochastic process with a PSD that rises at low frequencies is referred to as red noise. Since the GWB signal is common to all pulsars, this signal is said to produce a `common red noise'.
Given the capabilities of current PTA datasets, we assume that the common red noise is completely described by the parameters $A_\GWB$ and $\gamma_\GWB$.

In the presence of a GW signal, the angular correlation function is given by the so-called overlap reduction function \citep{AllenRomano_1999,Mingarelli_2013}
\begin{equation}\label{eq:Gamma}
    \Gamma_{ab} \propto \int_{S^2} \! \mathrm{d}\Hat{\Omega} \; \mathcal{P}(\Hat{\Omega}) \left[ \mathcal{F}_a^+(\Hat{\Omega})\mathcal{F}_b^+(\Hat{\Omega})
    +  \mathcal{F}_a^\times(\Hat{\Omega})\mathcal{F}_b^\times(\Hat{\Omega})\right], 
\end{equation}
where $\mathcal{P}(\Hat{\Omega})$ is the probability density function for GW power at different locations on the sky, $\hat{\Omega}$, as calculated from the Solar system barycentre, and 
\begin{equation}
    \mathcal{F}_a^A(\Hat{p}, \Hat{\Omega}) = \frac{1}{2}\frac{\Hat{p}_a^i\Hat{p}_a^j}{1-\Hat{\Omega}\cdot\Hat{p}_a}\, e_{ij}^A\left(\Hat{\Omega}\right) = \frac{1}{2}\frac{\Hat{p}_a^i\Hat{p}_a^j}{1+\Hat{k}\cdot\Hat{p}_a}\, e_{ij}^A\left(\Hat{\Omega}\right) 
\end{equation}
is the antenna factor for a GW with polarisation state $A$, propagating along the vector $\hat{k}$, as measured by a pulsar located in the direction $\Hat{p}_a$. The term $e^A_{ij}(\hat\Omega)$ denotes the $(i,j)$ components of the polarisation tensor for a GW with polarisation $A$, in the polarisation basis from, for example, \cite{Taylor_2020}. 

Our goal is to obtain an estimate for $\mathcal{P}(\Hat{\Omega})$. To this end we express $\mathcal{P}(\Hat{\Omega})$ in terms of a suitable basis, so that the basis coefficients can be constrained with PTA data. 
Following \cite{MPTA2025_aniso}, we employed a spherical harmonic basis,
\begin{equation}\label{eq:cSpH}
    \mathcal{P}(\Hat{\Omega}) = \sum_{l=0}^{\lmax}\sum_{m=-\ell}^{\ell} P_{\ell m} Y_{\ell m}(\Hat{\Omega}),
\end{equation}
with complex-valued spherical harmonics, $Y_{\ell m}(\Hat{\Omega})$ weighted by real-valued coefficients, $P_{\ell m}$ \citep{Cornish_2001, Kudoh_2005, Cornish2014}. Owing to the finite spatial resolution of a PTA, the sum truncates at an $\lmax$ value. The details of determining $\lmax$ are discussed in Sec.~\ref{sec:rev_sph_scheme}.

Using notation consistent with \cite{MPTA2025_aniso}, henceforth indices $\alpha$ and $\beta$ denote pulsar pairs $(ab)$; $\mu$ and $\nu$ denote spherical harmonics.
We write the overlap reduction function using the Einstein summation convention as
\begin{align}
    \Gamma_\alpha &= \int d\Omega \left[ \mathcal{F}^+_a(\hat\Omega) \mathcal{F}^+_b(\hat\Omega) + \mathcal{F}^\times_a(\hat\Omega) \mathcal{F}^\times_b(\hat\Omega)\right] Y_\mu(\hat\Omega) P_\mu \\
    &= R_{\alpha \mu} P_\mu ,
\end{align}
where $R_{\alpha\mu}$ is referred to as the `response matrix'.

\subsection{Per-frequency pulsar pair cross-correlations}

In the frequentist framework \citep{Mitra_2008,Thrane_2009}, the overlap reduction function is estimated from $N_\text{pairs}$ distinct pulsar pair cross-correlations, $\{\rho_\alpha\}$, designed such that \citep{Demorest_2013,Chamberlin_2015}
\begin{align}
\langle\rho_\alpha\rangle = \Gamma_\alpha A_\GWB^2 .
\end{align} 
Individual binaries are the most anticipated source of anisotropy \citep{SesanaVecchio_2010,Cornish2014,Burke-Spolaor_2019}. As a single binary emits GWs in a narrow frequency band, PTA anisotropy studies aim to calculate narrow-band sky maps at individual frequency bins of the PTA \citep{Pol_2022,Agazie_2023,MPTA2025_aniso}.

We calculated these maps from narrow-band pulsar pair cross-correlations, obtained from the narrow-band optimal statistic \citep{Gersbach_2025}, implemented in the \textsc{Python} package \textsc{defiant}.\footnote{This approach is mathematically equivalent to the calculations in \cite{MPTA2025_aniso}, but the code is simpler and streamlined with standard PTA GW searches thanks to the narrow-band optimal statistic. We verified consistency with the \cite{MPTA2025_aniso} results by comparing the numerical values of all matrices used during the sky map calculation.}
For the pulsar pair $\alpha$ at the $k^\mathrm{th}$ frequency bin, the narrow-band pulsar pair cross-correlation is
\begin{equation}\label{eq:PFOS_correlation}
    \rho_\alpha(f_k)\equiv \rho_{k,\alpha} = \frac{\Vec{X}_a^T\Tilde{\phi}(f_k)\Vec{X}_b}{\Tr\left[Z_a \Tilde{\phi}(f_k) Z_b \Phi(f_k) \right]},
\end{equation}
where 
\begin{align}
    \Vec{X}_a &= \mtx{F}^T_a \mtx{C}_a^{-1} \vct{\delta t}_a, \\
    \mtx{Z}_a &= \mtx{F}^T_a \mtx{C}_a^{-1} \mtx{F}_a.
\end{align}
The frequency selector matrix, 
\begin{equation}
    \Tilde{\phi}(f_k) = \mathrm{diag}(0,0,\ldots, \underbrace{1}_{2k-1},\underbrace{1}_{2k}, \ldots, 0,0),
\end{equation}
encodes the frequency bin at which the cross-correlation is evaluated\footnote{$\Tilde{\phi}(f_k)$ has the shape $(2k_\mathrm{max}\times 2k_\mathrm{max})$ and two adjacent non-zero components, because the Fourier transform matrices contains both a sine and a cosine component for each frequency.}.
The frequency-domain pulsar pair covariance matrix is normalised like so:
\begin{equation}
    \Phi(f_k) = \phi/S(f_k),
\end{equation}
ensuring that
\begin{equation}
    \langle\rho_{k,\alpha}\rangle  = \Gamma_\alpha S(f_k).
\end{equation}

\subsection{Sky map calculation in spherical harmonics decomposition}

With the set of cross correlations, $\{\rho_{k,\alpha}\}$, a GW sky map was obtained by calculating the coefficients, $\Vec{P}'$, that maximise the cross-correlation likelihood \citep{Mitra_2008,Thrane_2009, Pol_2022}. The equations governing the sky map calculation \citep[e.g.\ in][]{Pol_2022,MPTA2025_aniso} hold for both a single frequency as well as broadband pulsar pair correlations, so adopting the notation introduced in Eq.~\eqref{eq:PFOS_correlation}, the likelihood reads
\begin{equation}\label{eq:likelihood}
    {\cal L}\left(\Vec{\rho}_k | \Vec{P}_k\right) = \frac{\exp\left[-\frac{1}{2}\left(\Vec{\rho}_k - \mtx{R}\Vec{P}_k\right)^\dag \mtx{\Sigma}_k^{-1}\left(\Vec{\rho}_k - \mtx{R}\Vec{P}_k\right)\right]}{\sqrt{\det(2\pi\mtx{\Sigma}_k)}},
\end{equation}
where $\mtx{\Sigma}_k$ is the covariance matrix of the cross-correlations at the GW frequency $f_k$. We included off-diagonal terms in $\mtx{\Sigma}_k$ to take into account GW-induced pulsar-pair covariance \citep{Allen_2023_HD71}. 
The detailed expressions for the full matrix entries can be found in the appendices in \cite{Gersbach_2025} and \cite{MPTA2025_aniso}. 

The maximum likelihood estimator is given as \citep{Mitra_2008,Thrane_2009,Pol_2022,MPTA2025_aniso}
\begin{equation}\label{eq:ml_estimator}
    \Vec{P}'_k = \mtx{M}_k^{-1} \Vec{X}_k ,
\end{equation}
where the vector
\begin{equation}
\Vec{X}_k \equiv \Vec{X}(f_k)= \mtx{R}^\dag \mtx{\Sigma}_k^{-1} \, \Vec{\rho}_k  
\label{eq:dirty_map}
\end{equation}
is known as the dirty map, i.e. an inverse-noise weighted representation of the GW power on the sky as seen through the response of the pulsars \citep{Thrane_2009,Pol_2022,Agazie_2023,Ali-Haimoud_2021}.
Meanwhile,
\begin{equation}
\mtx{M}_k \equiv \mtx{M}(f_k) = \mtx{R}^\dag\mtx{\Sigma}_k^{-1}\mtx{R} 
\label{eq:Fisher_matrix}
\end{equation}
is the Fisher matrix of the maximum likelihood estimators.

Following the findings and arguments in \cite{MPTA2025_aniso}, we regularised the Fisher matrix inversion in Eq.~\eqref{eq:ml_estimator}. This was done with the singular value decomposition scheme outlined in \cite{Thrane_2009,Abadie_2011,MPTA2025_aniso}, whereby the first $\sreg$ sorted eigenvalues of $\mtx{M}$ were kept, and beyond that threshold, the remaining smaller ones were set to infinity\footnote{This means that for a spherical harmonics expansion to $\lmax$, where we have $(\lmax+1)^2$ modes, $\sreg\leq (\lmax+1)^2$.}. This accounts for the variations in the constraints of the different eigenmodes of $\mtx{M}$ due to the sky distribution and sensitivity of the pulsars. The regularised maximum likelihood estimate is 
\begin{equation}\label{eq:reg_ml_estimator}
    \Tilde{\Vec{P}}'_k = \Tilde{\mtx{M}}_k^{-1} \Vec{X}_k,
\end{equation}
where $\Tilde{\mtx{M}}_k^{-1}$ is the pseudo-inverse of $\mtx{M}_k$ containing only the first $\sreg$ eigenmodes of $\mtx{M}_k$. The tildes denote regularised quantities. Choosing the number of modes to keep is a trade-off. By throwing out some modes, the sky map becomes biased because these rejected modes are no longer included in the sky maps. But on the other hand, we also remove poorly constrained modes that contribute only noise to the sky map.

In contrast to the dirty map, the maximum-likelihood estimators $\Vec{P}'_k$ and $\Tilde{\Vec{P}}'_k$ (Eqns.~\eqref{eq:ml_estimator},\eqref{eq:reg_ml_estimator}) are referred to as the clean map\footnote{As is explained in more detail in \cite{MPTA2025_aniso}, $\Vec{P}'_k$ and $\Tilde{\Vec{P}}'_k$ can have negative values, due to residual noise present in the data.}. So in summary, calculating a frequentist GW sky map starts with obtaining the dirty map from the pulsar pair correlations and the Fisher matrix from the PTA noise models. Then, the regularised inverse of the Fisher matrix is calculated, and multiplied with the dirty map to obtain the regularised clean map.

The uncertainty associated with this regularised clean map is
\begin{equation}
    \tilde{\sigma}^{\tilde{P}'_k}_\mu = \sqrt{ {\mtx{\Tilde{M}}_k^{-1}}_{\mu\mu}}.\label{eq:var_power_sph}
\end{equation}
The clean map signal-to-noise ratio (S/N) is given by 
\begin{align}\label{eq:clean_snr}
    S/N = \frac{\tilde{\mathcal{P}}'_{k,\hat\Omega}}{\tilde{\sigma}_{\hat{\Omega}}^{\tilde{P}'_k} }.
\end{align}
Here the subscript $\hat\Omega$ indicate that we have transformed the numerator and denominator from the spherical harmonic basis to the pixel basis. 

In the following, we drop the subscript $k$ and implicitly assume frequency resolved quantities. All other previously introduced conventions and notations remain unchanged.

\section{Sky resolution of a PTA dataset}
\label{sec:sky_resolution}

We demonstrated in our previous work that the sky map output from a PTA dataset depends on the data analysis scheme. Hence, a quantity measuring the resolving properties should reflect all aspects of the analysis, especially the influence of the regularisation scheme. 
Due to the anisotropic distribution of pulsars on the sky, it is natural to expect that the resolvability of a PTA varies with sky position; in regions with a high sky density of pulsars, we expect the PTA to resolve structure better than in regions sparsely populated with pulsars.

\subsection{PTA point spread function}
\label{ssec:PSF_scheme}

The resolution of an imaging tool can be characterised by its ability to resolve point sources, captured in the so-called PSF. In the context of GW astronomy, this corresponds to determining how a single GW point source at the sky position $\Hat{\Omega}_i$ appears on the clean sky map after observing it with the method described in the previous section.

First, we have to choose a suitable representation of the point source.
For a visualisation on a pixellated sky map, we transformed the spherical harmonics maximum likelihood solution, $\Vec{P}$, into its pixel basis representation, $\Vec{P}_{\hat{\Omega}}$, via
\begin{align}
\Vec{P} = U^\dagger\Vec{P}_{\hat{\Omega}}, 
\end{align}
where $\mtx{U}$ is a unitary basis transformation matrix. The centre of each pixel represents an (RA, DEC) co-ordinate pair, and the pixel size was chosen such that the smooth features of the spherical harmonics representation are properly displayed on the resulting sky map. A side effect of this fine-grained pixelation is that the pixel size is well below the average angular separation of two pulsars. Hence, we can model a GW point source located at $\Omega_i$ with a single `hot' pixel at that position. 
Using vector notation, a single hot pixel at the i$^{th}$ pixel position can be written as 
\begin{equation}
    \Vec{P}_{\hat{\Omega}}(\Hat{\Omega}_i) \equiv \Vec{P}_{\hat{\Omega},i} = (0, \ldots, \underbrace{1}_{i^\mathrm{th} \text{entry}}, \ldots,0)^T.
\end{equation}

Capturing this hot pixel through the PTA leads to the dirty map
\begin{equation}
    \Vec{X}_i = \mtx{M} \mtx{U}^\dagger \Vec{P}_{\hat{\Omega},i},
\end{equation}
where we have used Eq.~\eqref{eq:ml_estimator}, explicitly including the basis transformation.

In the usual anisotropy analysis, this dirty map is deconvolved with the regularised dirty map, $\Tilde{M}$, to obtain the clean map. Hence, the full expression for the clean map representation of the hot pixel is given as
\begin{equation}\label{eq:hot_pixel_recovery}
   \Vec{P}_{\hat{\Omega},i}' = \mtx{U}\Tilde{\mtx{M}}^{-1} \Vec{X}_i = \mtx{U}\Tilde{\mtx{M}}^{-1} \mtx{M} \mtx{U}^\dagger \Vec{P}_{\hat{\Omega},i}.
\end{equation}
This equation describes the recovered clean map under the assumption that the pulsar noise models are a perfect representation of the actual pulsar noise budget, i.e.\ $\Tilde{\mtx{M}}^{-1}$ and $\mtx{M}$ contain the same noise values. In real PTA datasets, the clean map representation of a GW point source can be influenced by noise misspecification. For completeness, we provide a derivation and analysis of the full expression for the recovered clean map in App.~\ref{app:generalised_PSF_formalism}, which also accounts for an (assumed) noise misspecification. We also demonstrate that it recovers Eq.~\eqref{eq:hot_pixel_recovery} in the limiting case of perfect pulsar noise models.

\begin{figure}
    \centering
    \includegraphics[width=0.9\linewidth]{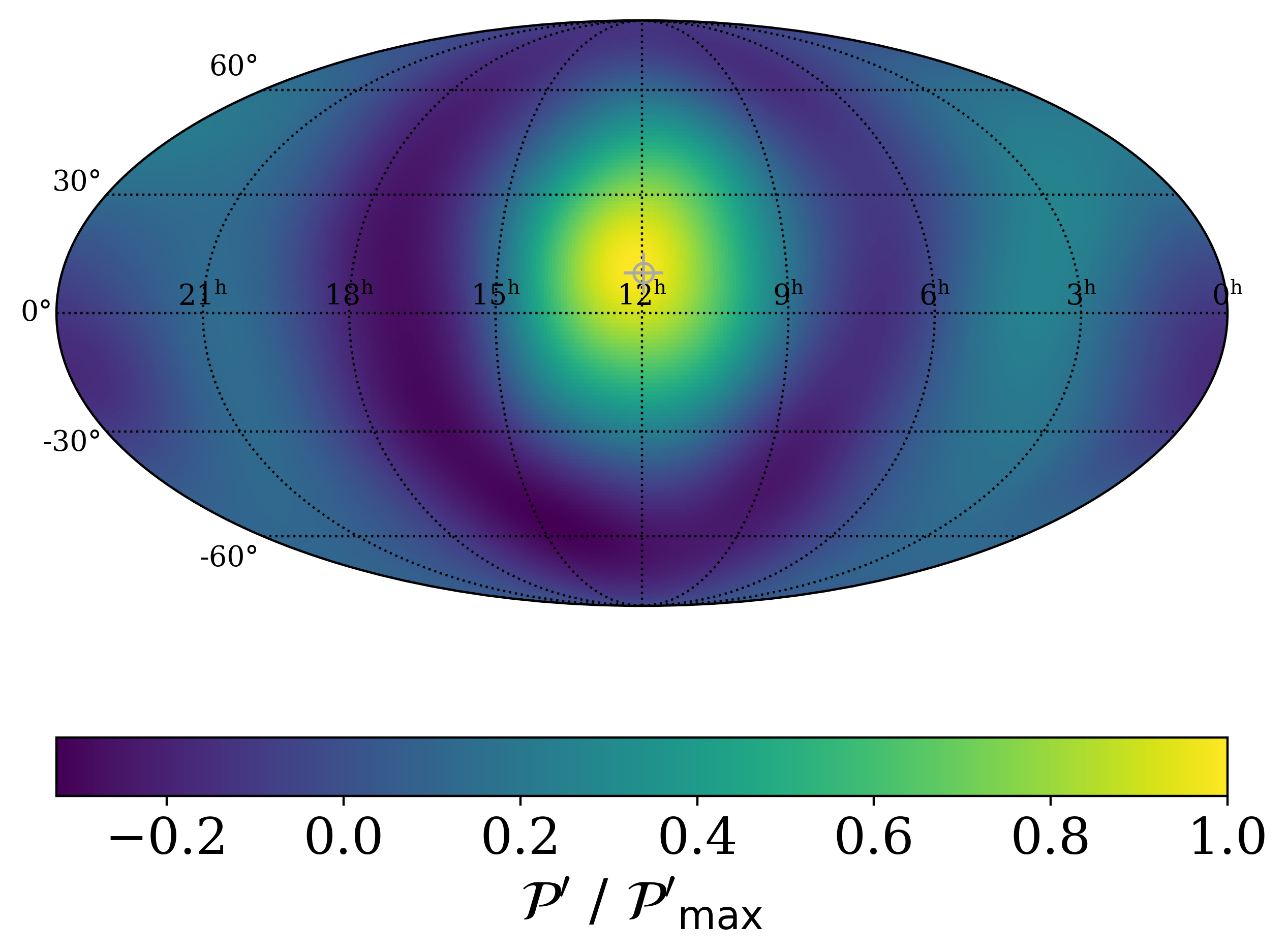}
    \includegraphics[width=0.9\linewidth]{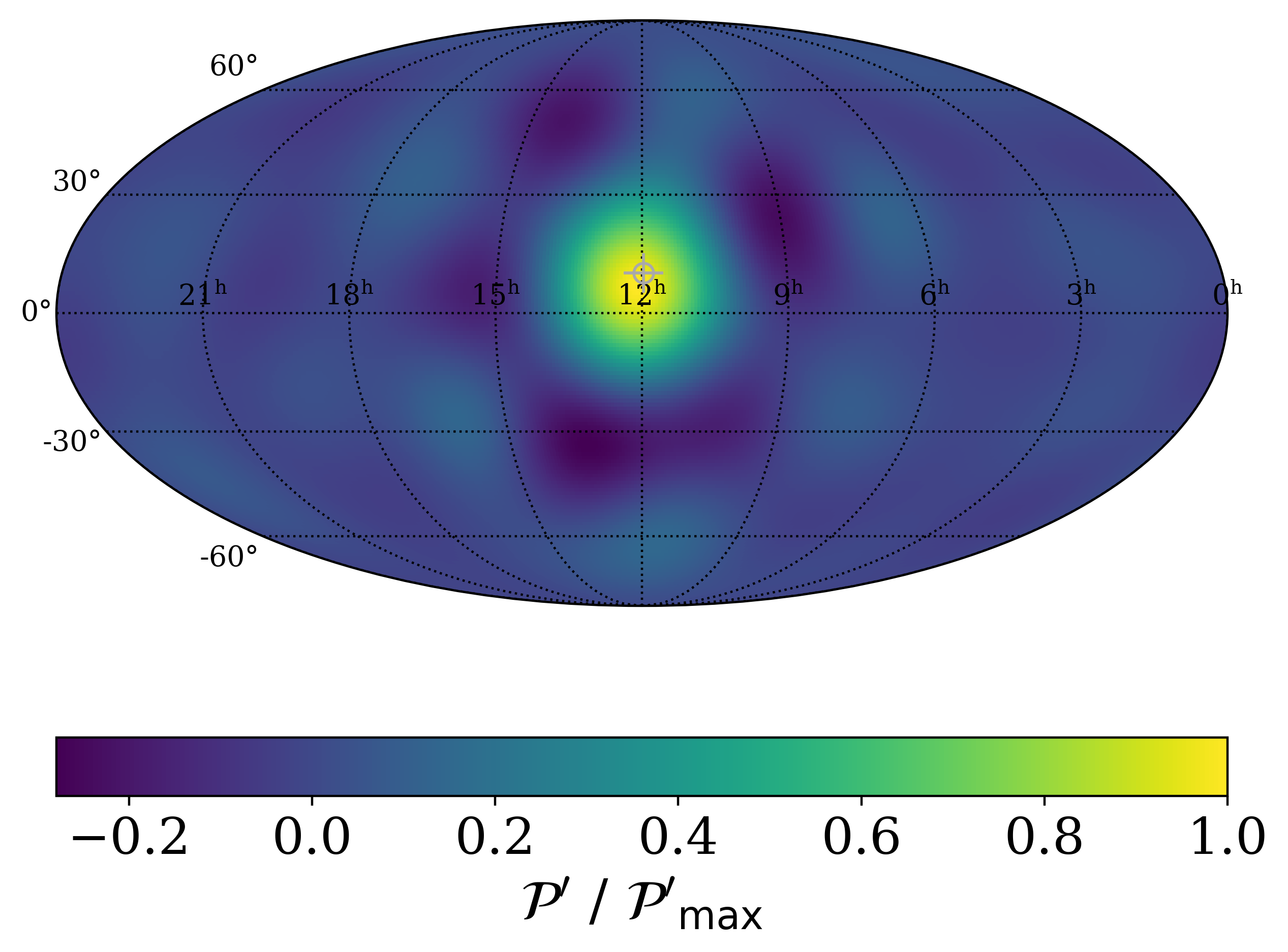}
    \caption{Single hot pixel seen through a 40PSR PTA with different regularisation cut-offs. The crosshair indicates the position of the hot pixel. Top: Clean map with $\lmax=8$, $s_\mathrm{reg}=15$. Bottom: Clean map with $\lmax=8$, $s_\mathrm{reg}=40$.}
    \label{fig:testpixel_40PSR}
\end{figure}

In Fig.~\ref{fig:testpixel_40PSR} we show the transformation process due to the regularisation using a single hot pixel observed through a simple PTA consisting of 40 pulsars equally distributed on the sphere using the Fibonacci lattice technique.\footnote{For $N$ points on the sphere, their positions are given by $\theta_i = (2\pi i)/\phi$ and $\varphi = \arccos(1-2i/N)$, with the golden ratio $\phi= \left(1+\sqrt{5}\right)/2$.} We can clearly see the blurring due to the regularisation taking place. 

From this clean sky map, we can now calculate the value of the PSF at the $i^\mathrm{th}$ pixel, $A_{\mathrm{PSF},i}$, by determining the area to which a hot pixel is blurred out as 
\begin{equation}
    A_{\mathrm{PSF},i} = A\left(\Vec{P}_{\hat{\Omega}}' > 0.5~\Vec{P}_{\hat{\Omega},\mathrm{max}}'\right),
\end{equation}
where we defined the blurred area to include all pixels with values larger than half the maximum of the sky map.

In practice, $A_{\mathrm{PSF},i}$ is calculated by normalising $\Vec{P}_{\hat{\Omega},i}'$ by its maximum value, and adding up the areas of all pixels whose clean map values are larger than $0.5$. The area of each pixel is obtained from the small area approximation on the sphere, ensuring that the sum of all pixel areas add up to $4\pi$.

If this evaluation of the point spread area is done iteratively for all pixels, we obtain the PSF sky map, i.e.\ the point spread area as function of sky position. The values of the PSF map have units of areas.

Notably, this method allows for the addition of disjoint areas, as long as their PSF values are above the threshold. This, in turn, can lead to unreliable PSF estimates. Possible solutions could include an alternative effective-area definition, or imposing a connected-area condition on the PSF estimation. 
However, we determine that this failure mode happens rarely and only in areas with poor resolution where the GW power of the point source is spread out widely across the map, making the map maximum barely distinguishable from the noise floor.
In areas with a reasonably high resolution, this issue does not arise. As the following work focuses on the sky areas in which the PTA has the best resolution, we defer mitigating this issue to a future work. 

We note that in general a PSF features both an area as well as a location of a recovered point source. Investigating sky maps similar to Fig.~\ref{fig:testpixel_40PSR}, we also found that the sky position of the clean $S/N$ map maximum does not always correspond to the sky position of injected the hot pixel. This behaviour is likely caused by superposition of antenna patterns from the pulsars. In areas densely populated by pulsars, there is little to no position offset along the connecting lines of next-neighbour pulsars, but a slight dislocation if the source is encircled by pulsars. This offset does not go beyond the area that is enclosed by the pulsars, and also falls into the PSF extent. For sources that are in an area with little to no pulsar population on a large scale, the S/N recovered from the source is close to the noise floor. This leads to large displacements, but in turn they are not associated with a source detection. So the PTA geometry and analysis pipeline can also introduce a mismatch between the actual and recovered source position. While this position distortion feature is of interest for identifying the host of a single supermassive black hole binary GW signals, and could be investigated further, for instance in the scope of a hotspot-galaxy-catalogue-crossmatching study, our focus remains on investigating the PSF area behaviour.

\subsection{The PTA distortion matrix}
\label{ssec:distortion_matrix}

In order to find an efficient implementation of the PSF calculation scheme, we return to Eq.~\eqref{eq:hot_pixel_recovery}. Looking closer, we find that the multiplication with $ \Vec{P}_{\hat{\Omega},i}$ is nothing but selecting the $i^\mathrm{th}$ column of the matrix product $\mtx{U}\Tilde{\mtx{M}}^{-1} \mtx{M} \mtx{U}^\dagger$, as $\Vec{P}_{\hat{\Omega},i}$ can be interpreted as the canonical basis vectors of a $N_\mathrm{pix}$-dimensional Euclidean space. Conveniently, all quantities in Eq.~\eqref{eq:distortion_matrix} can be accessed as by-products of the existing sky map analysis framework. So, the calculation of the PSF due to the regularisation scheme becomes an iterative evaluation of the column vectors of $\mtx{U}\Tilde{\mtx{M}}^{-1} \mtx{M} \mtx{U}^\dagger$.

Hence, this matrix holds all hot-pixel clean sky maps as its column vectors, and hence all information about the PSF. This motivates the defining of this matrix product as the $N_\mathrm{pix}\times N_\mathrm{pix}$ `distortion matrix' of a PTA with a regularisation cut-off, $\sreg$, 
\begin{equation}\label{eq:distortion_matrix}
    \mtx{\Lambda}^\mathrm{pix}_{\sreg} := \mtx{U}\Tilde{\mtx{M}}^{-1} \mtx{M} \mtx{U}^\dagger = \mtx{U}\mtx{\Lambda}_{\sreg} \mtx{U}^\dagger.
\end{equation}
The quantity $\mtx{\Lambda}_{\sreg}$ is its spherical harmonics-space representation. Intuitively, $\mtx{\Lambda}^\mathrm{pix}_{\sreg}$ (or $\mtx{\Lambda}_{\sreg}$) contains all information about the smear that the PTA introduces to point-like GW sources, due to the regularisation of the Fisher matrix inversion. 

Comparing Eq.~\eqref{eq:distortion_matrix} to the definition of the PTA clean sensitivity map,  Eqns.~(D2) and (D3) in \cite{MPTA2025_aniso}, we find that the sensitivity map is calculated from the main diagonal of $\mtx{\Lambda}_{\sreg}^\mathrm{pix}$. This is not surprising, as both quantities share the underlying question of how a point source appears in the analysis product. The mathematical similarity demonstrates not only the closeness between these two methods, but also the importance of $\mtx{\Lambda}_{\sreg}^\mathrm{pix}$ and $\mtx{\Lambda}_{\sreg}$, respectively, in the context of characterising a PTA.

Investigating $\mtx{\Lambda}_{\sreg}$ from an algebraic point of view, we notice the following: rewriting it in terms of the singular value decomposition of $\mtx{M}$, $\mtx{M} = \mtx{V}\mtx{\Xi}\mtx{V}^\dagger$ with the unitary matrix $\mtx{V}$, we arrive at
\begin{align}\label{eq:lambda_def}
    \mtx{\Lambda}_{\sreg} := \Tilde{\mtx{M}}^{-1} \mtx{M} &= \mtx{V}\mtx{\Xi}^+\mtx{V}^\dagger \,\mtx{V}\mtx{\Xi}\mtx{V}^\dagger  \\
    & = \mtx{V} \mtx{\Delta} \mtx{V}^\dagger,
\end{align}
where $\mtx{\Delta} \equiv \mathrm{diag}(\underbrace{1,\ldots,1}_{\sreg},\underbrace{0,\ldots, 0}_{n-\sreg}) $.
Hence, $\mtx{\Lambda}_{\sreg}$ is the orthogonal projector onto the regularised subspace, i.e. the subspace spanned by the first $\sreg$ left singular vectors of $\mtx{M}$. 
The quantity $\mtx{\Lambda}_{\sreg}^\mathrm{pix}$ is the representation of this spectral projector in the pixel basis. As a consequence, the clean sky maps corresponding to the individual pixels are the column vectors of the spectral projector in the pixel basis. This is mathematically capturing the impact of the information loss due to the regularisation, required from the noise and geometry of the PTA.

\section{Revisiting the current spherical harmonics scheme}
\label{sec:rev_sph_scheme}
The number of spherical harmonic modes, $\{(\ell,m)\}$, that can be uniquely constrained cannot be larger than the number of pulsars. In the scheme currently employed for PTA sky map analyses \citep{Pol_2022,Agazie_2023,MPTA2025_aniso}, this is enforced by choosing \ $\lmax = \lfloor \sqrt{N_\mathrm{psr}}-1\rfloor$. We refer to this as the `classic scheme'.
The disadvantage of this approach is that it does not take into account the fact that the resolution of a PTA often varies across the sky.
We use the PSF maps developed in the previous section in order to allow for an adaptive scheme in which the resolution varies across the sky.

\subsection{Shortcomings of the currently used (classic) scheme}
\label{ssec:shortcomings}

For this investigation we simulated an artificial PTA consisting of 40 pulsars, isotropically distributed according to a Fibonacci lattice. All pulsars have a root mean square of \SI{100}{\nano\second}, regular ToAs over a total observation time span of \SI{10}{\yr}, and are simulated with white noise only. The analysed ToAs were simulated using \textsc{libstempo}.

\begin{figure}
    \centering
    \includegraphics[width=0.9\linewidth]{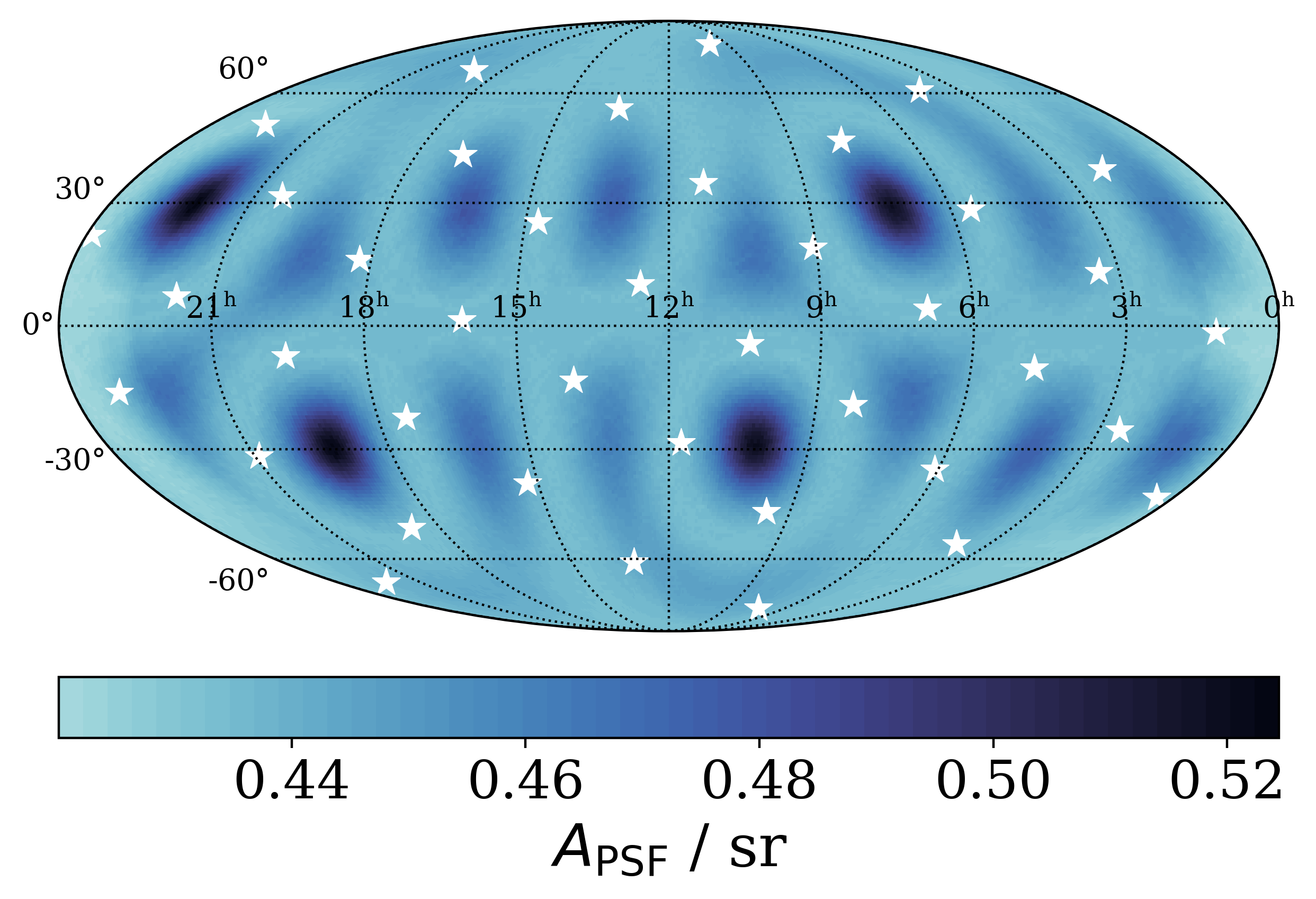}
    \includegraphics[width=0.9\linewidth]{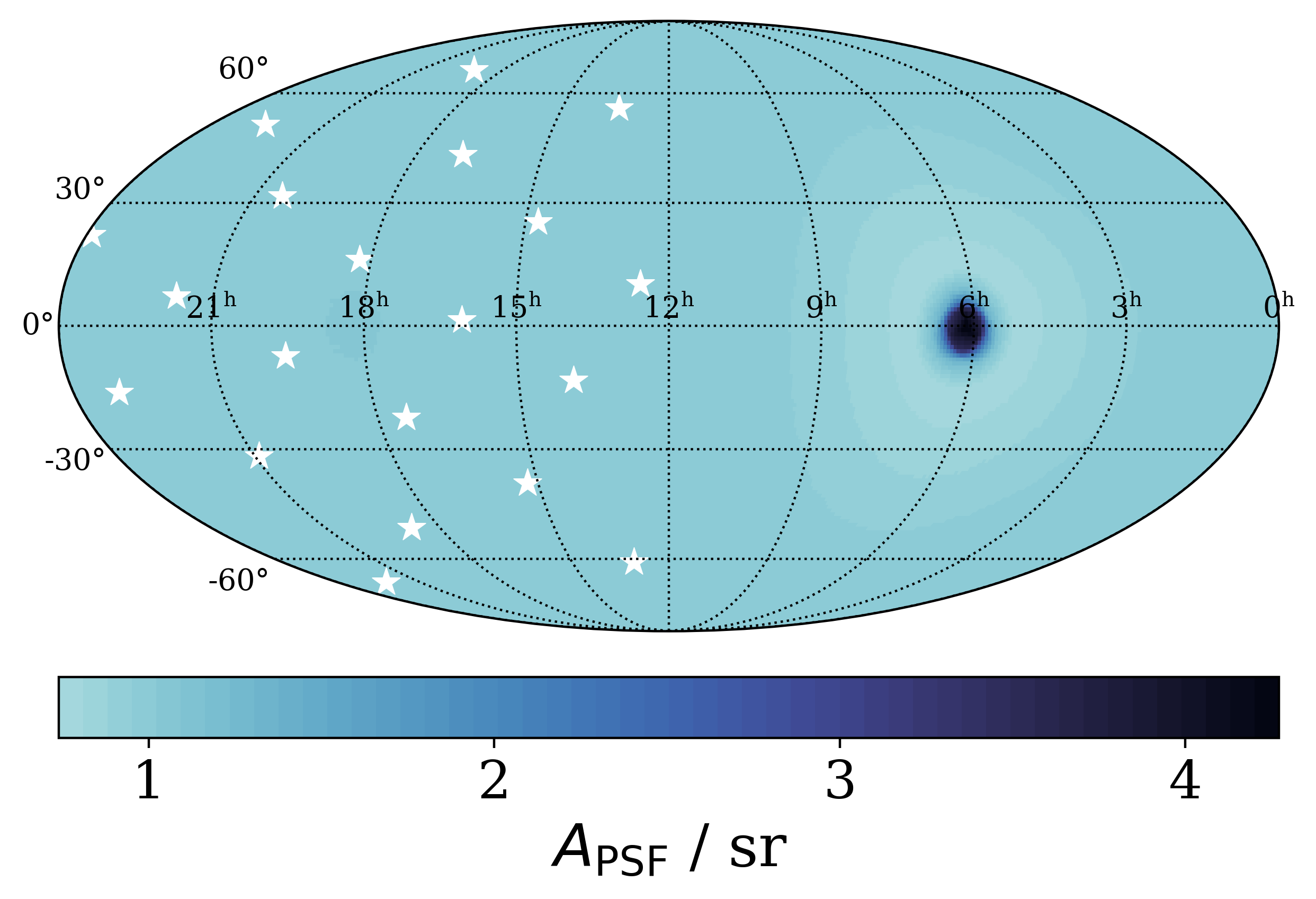}
    \caption{Sky maps showing the (S/N-based) PSF of the PTA as a function of sky position. The white stars indicate the positions of the pulsars in the respective PTA. Both plots were calculated using the classic spherical harmonics expansion scheme based on the number of pulsars. Top: 40 pulsar PTA, $\lmax=5, \sreg=35$. Bottom: 20 pulsar subset PTA, $\lmax=3, \sreg=15$.}
    \label{fig:test_ptas_currentscheme}
\end{figure}

One expects that such a PTA should allow for 40 equally constrained spherical harmonics modes, since all pulsars contribute equally in this PTA, and they are isotropically distributed across the sky.
But the classic scheme demands that $\lmax = 5$, so the spherical harmonics expansion contains only 36 modes. Following the scheme leads to purposefully ignoring a subset of the constraints provided by the PTA dataset, although all of them are equivalent. One major consequence of this negligence becomes imminently accessible from the PSF map of this set-up (40 pulsars, $\lmax = 5, \sreg=35$), shown in the top plot in Fig.~\ref{fig:test_ptas_currentscheme}. We would expect any pattern visible on the map to follow the position of the pulsars, as well as similar pulsar constellations to lead to similar PSF values. Yet, we find an irregular pattern that is not correlated with the pulsar positions. This is not expected and indicative of a sub-optimal set-up.

In the next step, we removed half of the pulsars in the 40 pulsar isotropic array with RA $< \SI{12}{\hr}$. In the classic scheme, this would allow for a spherical harmonics expansion to $\lmax = 3$. The corresponding PSF map of this set-up (20 pulsars, $\lmax = 3, \sreg=15$) is shown in the lower plot of Fig.~\ref{fig:test_ptas_currentscheme}. The increased values of the PSF show that while we only reduced the number of pulsars in one particular part of the sky, we gave up on spatial resolution across the whole sky. Moreover, we would logically expect the resolution of the PTA on the part of the sky where the pulsars were not removed to be the same as for the full 40 pulsar PTA. As we can see from the PSF map, this is clearly not the case. 

Thus, we have identified several shortcomings of the scheme currently used to set up spherical-harmonics-based anisotropy analyses. In the next step, we aim to mitigate these problems with a revised set-up procedure.

\subsection{An enhanced scheme for regularised gravitational-wave sky map calculation}

It is clear that the current scheme of choosing $\lmax$ based on the number of pulsars neither reflects the actual PTA sky resolution nor is flexible enough to adapt to uneven pulsar constellations. Hence we propose a shift in the paradigm. We chose the maximum spherical harmonics degree, $\lmax$, such that it reflects the pulsar distribution in the highest populated sky areas. The surplus modes, i.e.\ the over-resolution in sparsely populated sky areas, can then be removed using the established regularisation scheme.

The choice of $s_\mathrm{reg}$ is a trade-off between including additional unspecified noise and rejecting signal information. It reflects the difference in the sensitivity of the individual pulsars, but also the lack of support of those spherical harmonics modes falling in sky areas with a local underdensity in the pulsar distribution. In a sense, the surplus modes are no different to those modes already regularised in the classic scheme. None of them is well constrained. With this approach, the condition that no more modes than the number of pulsars can be constrained translates into a limit on the regularisation cut-off, namely $\sreg \leq N_\mathrm{PSR}$.

\subsubsection{Creating credible PSF sky maps}

\begin{figure*}
    \centering
    \includegraphics[width=0.24\textwidth]{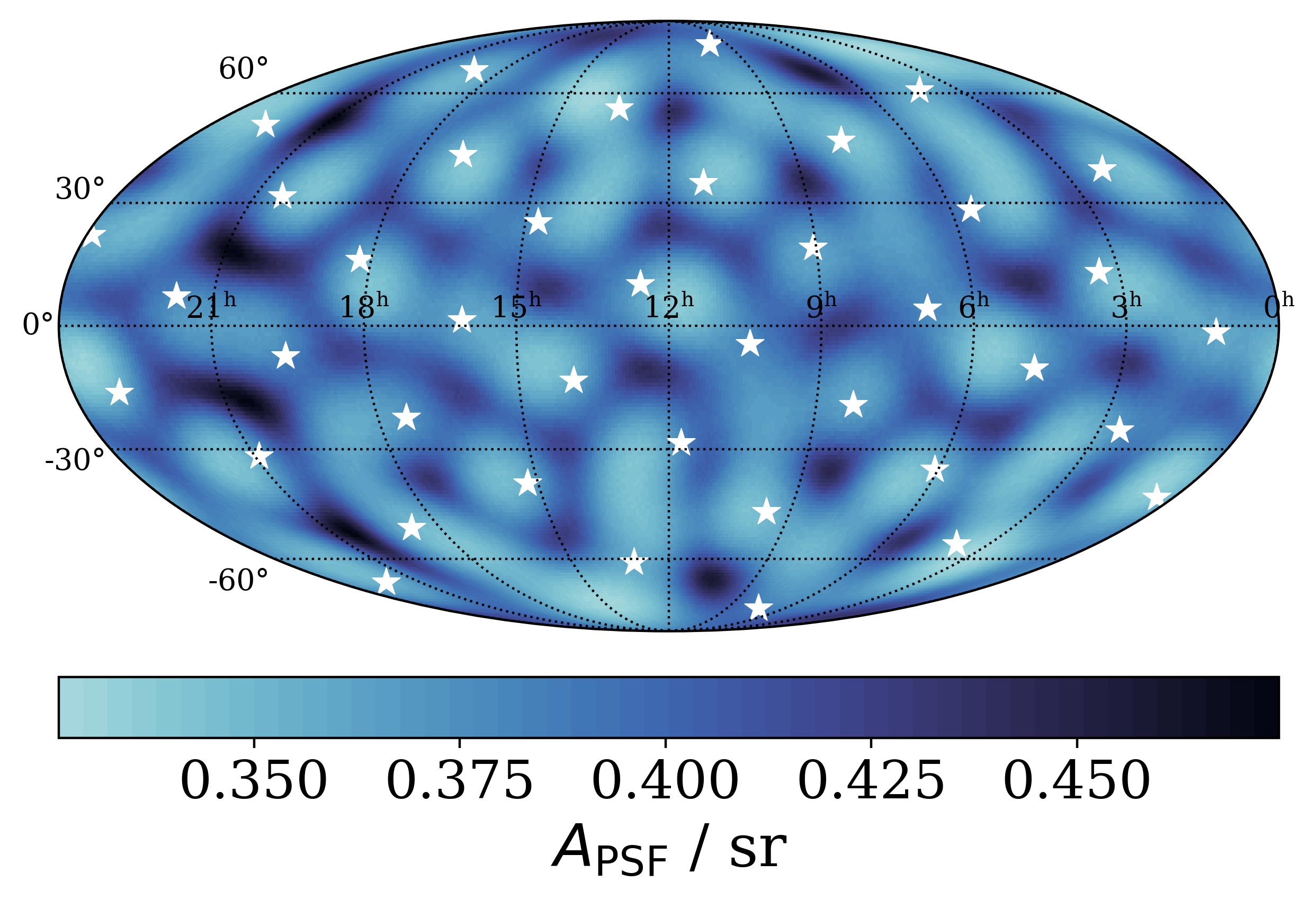}
    \includegraphics[width=0.24\textwidth]{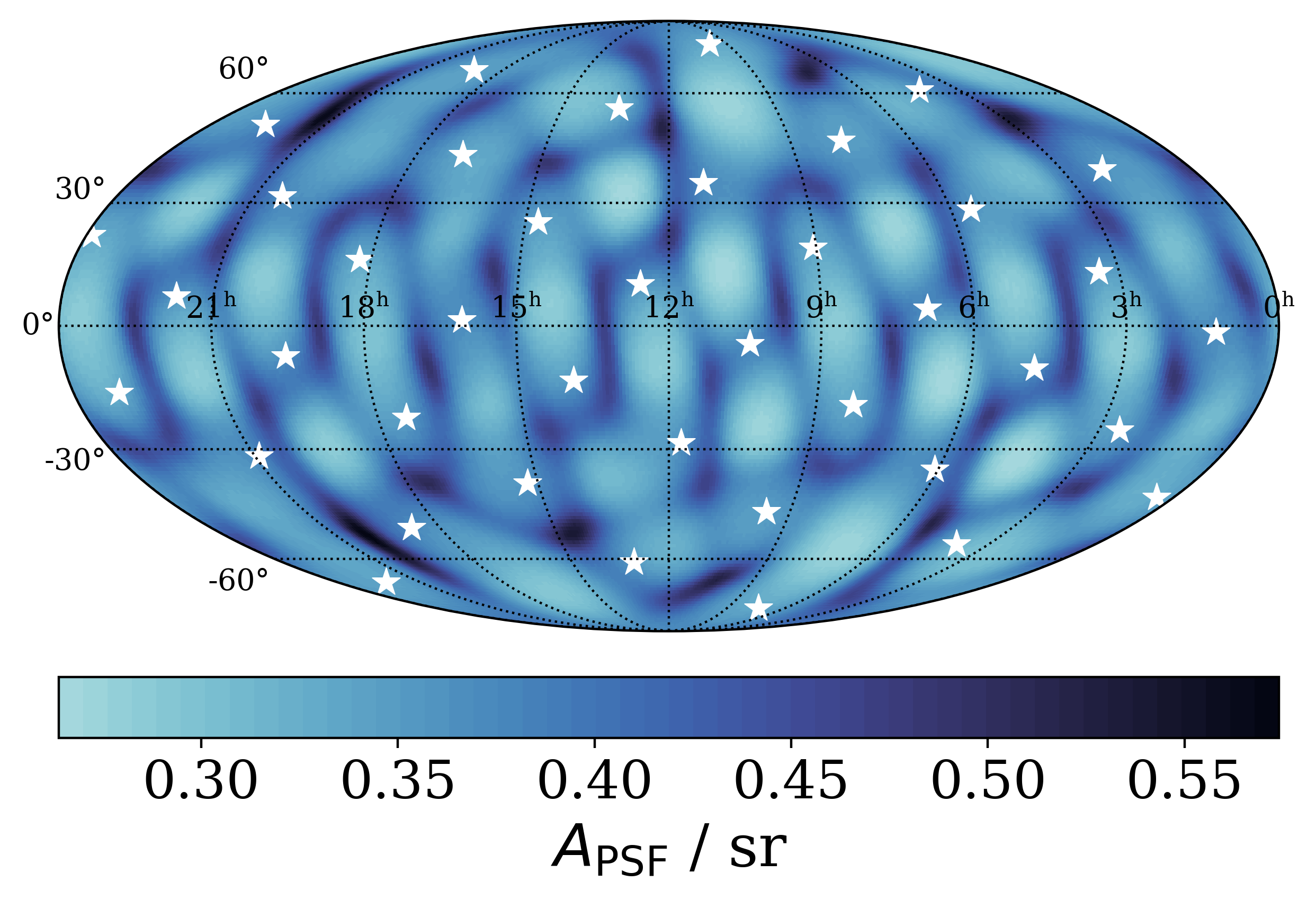}
    \includegraphics[width=0.24\textwidth]{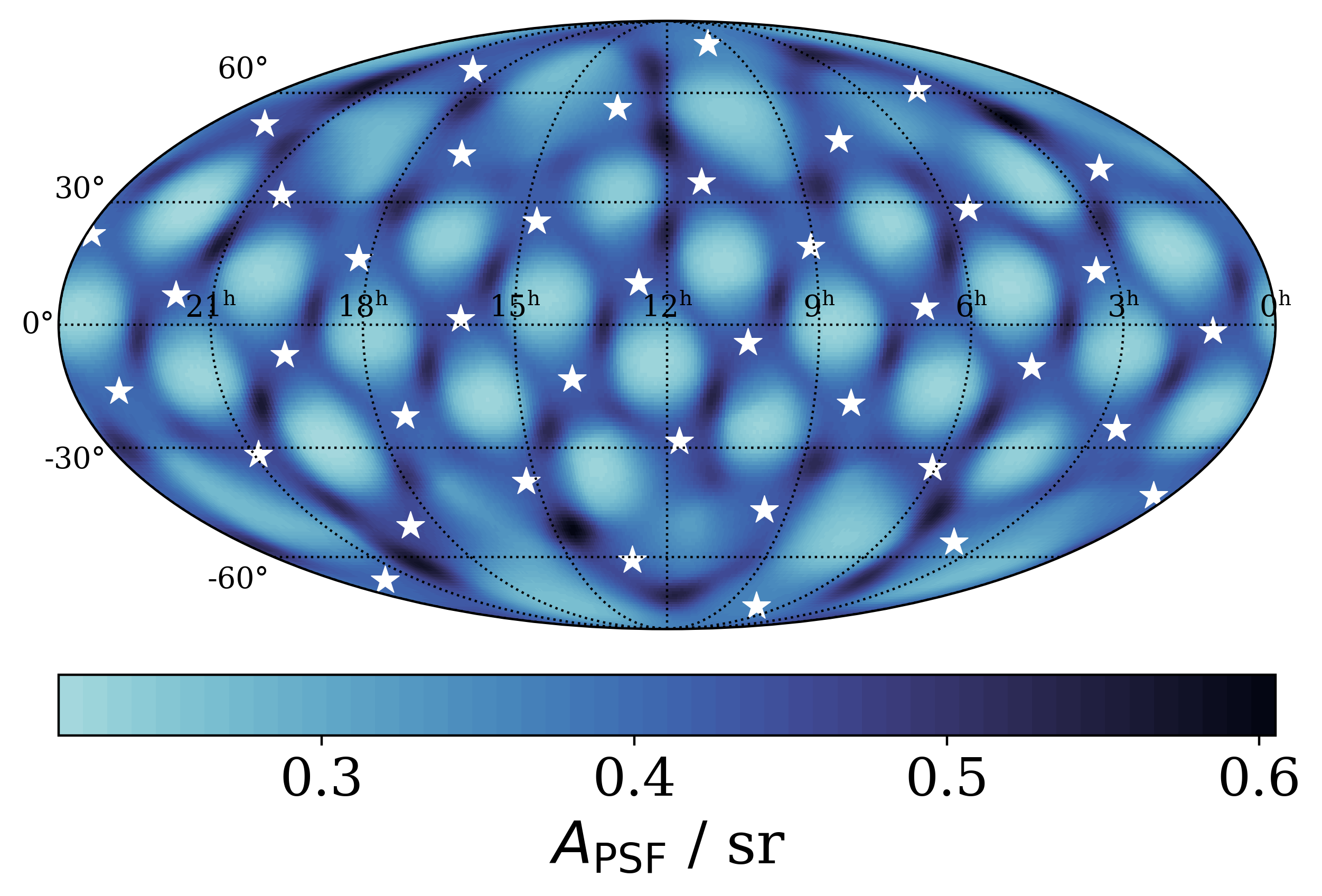}
    \includegraphics[width=0.24\textwidth]{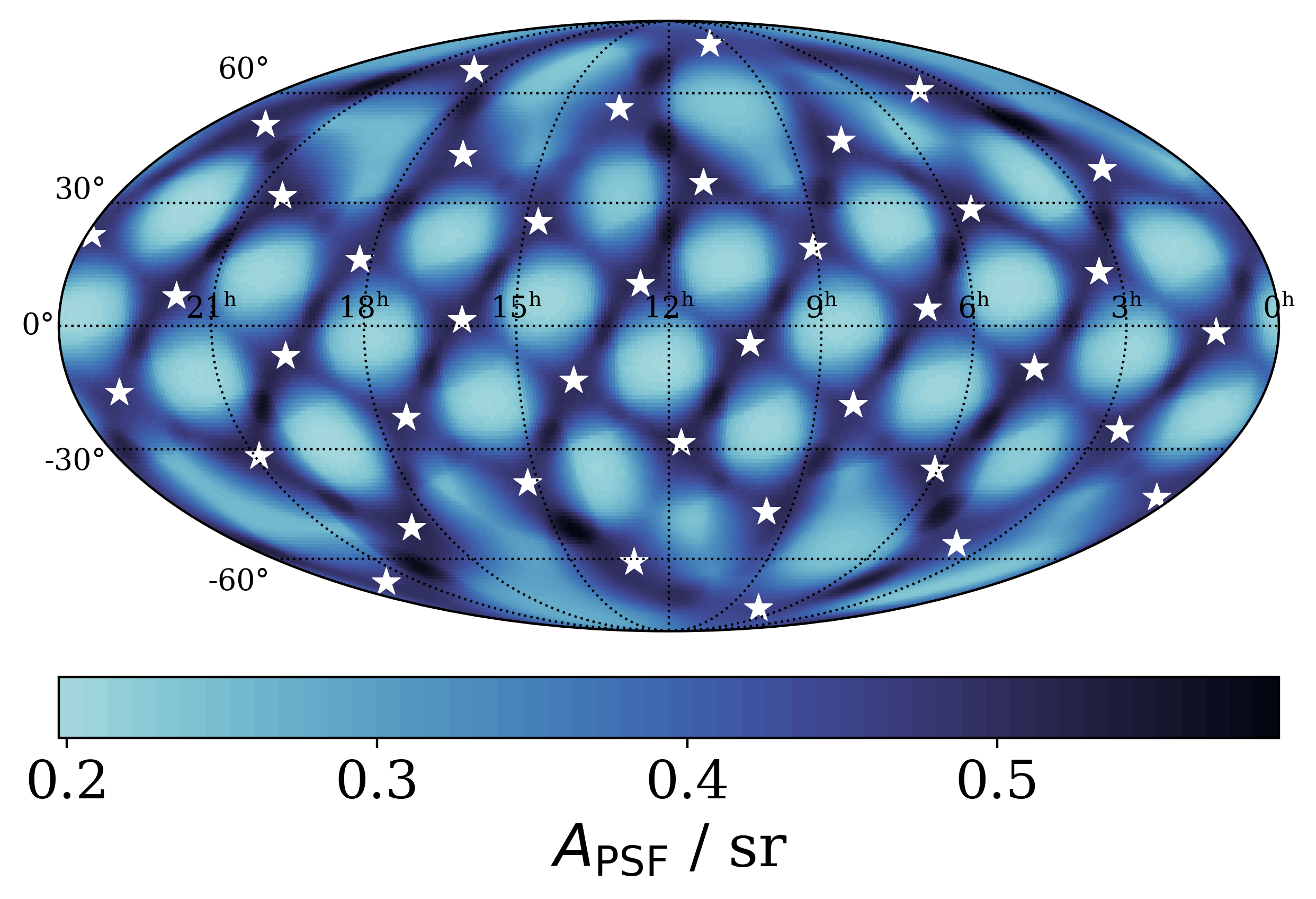}
    \caption{Point spread area sky maps for the 40 pulsar isotropic PTA set-up. From left to right: $\lmax = \{6,8,10,12\}$. The white stars indicate the pulsar positions.
    }
    \label{fig:40PSR_lmax-variation}
\end{figure*}

We explore the effect of increasing $\lmax$ with the full 40 pulsar PTA. Due to the symmetries in the set-up, we set $\sreg = 40$ and increased $\lmax$. The resulting PSF maps are shown in Fig.~\ref{fig:40PSR_lmax-variation}. The improvement in both the minimum PSF values as well as the regularity of the pattern across the sky is directly visible. 

For the larger values of $\lmax$, we find that the pattern reflects the expectation from the single pulsar response \citep{RomanoAllen_FAQ_2024}. The PSF is smaller in regions between four pulsars, but on connection lines between two adjacent pulsars it is larger, as only two pulsars constrain the source position.

This demonstrates that this approach allows us to model the PTA geometry more accurately. Moreover, we find that for $\lmax=\{10,12\}$ neither the range of the PSF values (range of the colour bar) nor the pattern on the map significantly changes. This is a first indication of some kind of convergence behaviour that can help to identify a suitable $\lmax$.

\begin{figure*}
    \centering
    \includegraphics[width=0.24\textwidth]{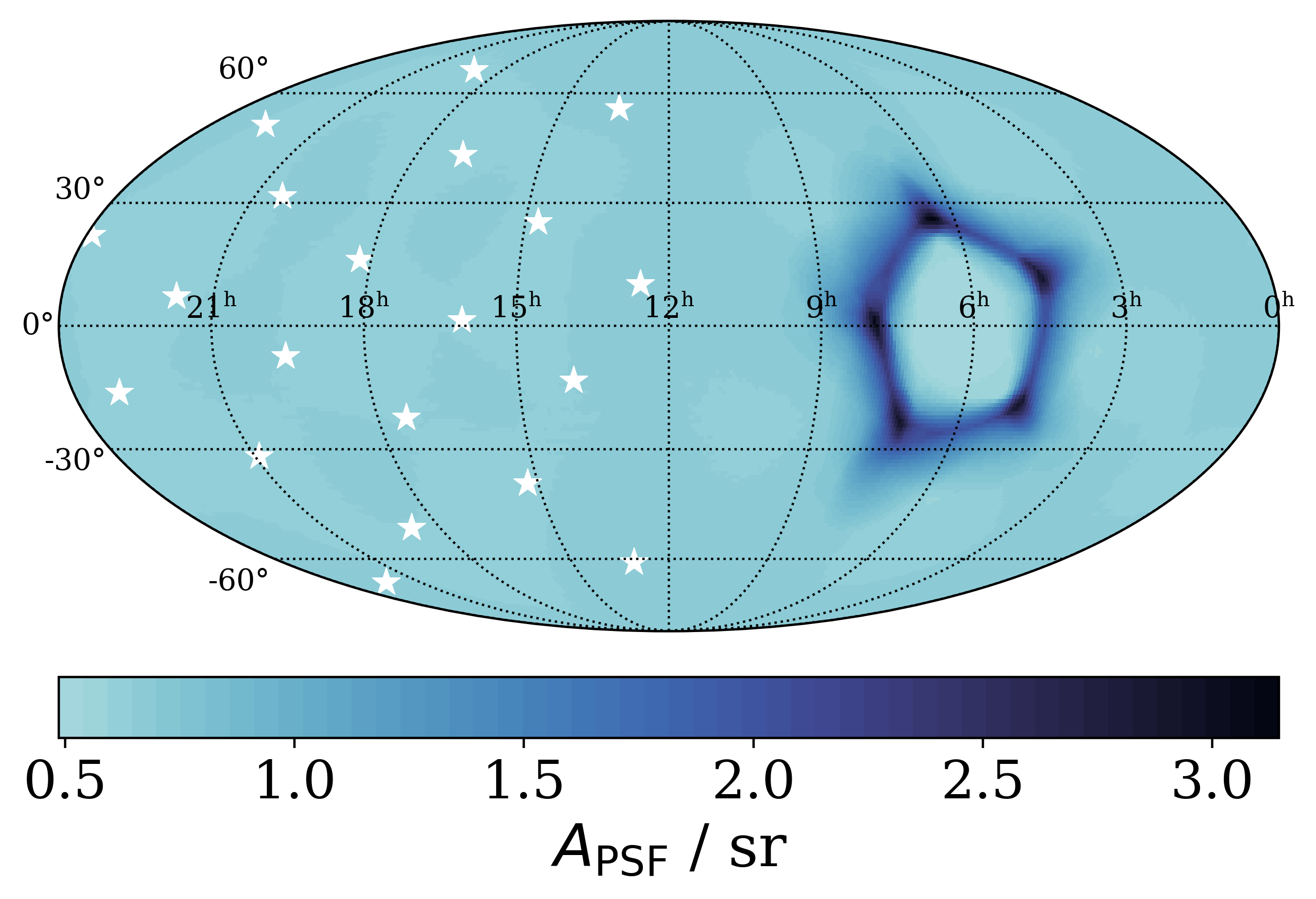}
    \includegraphics[width=0.24\textwidth]{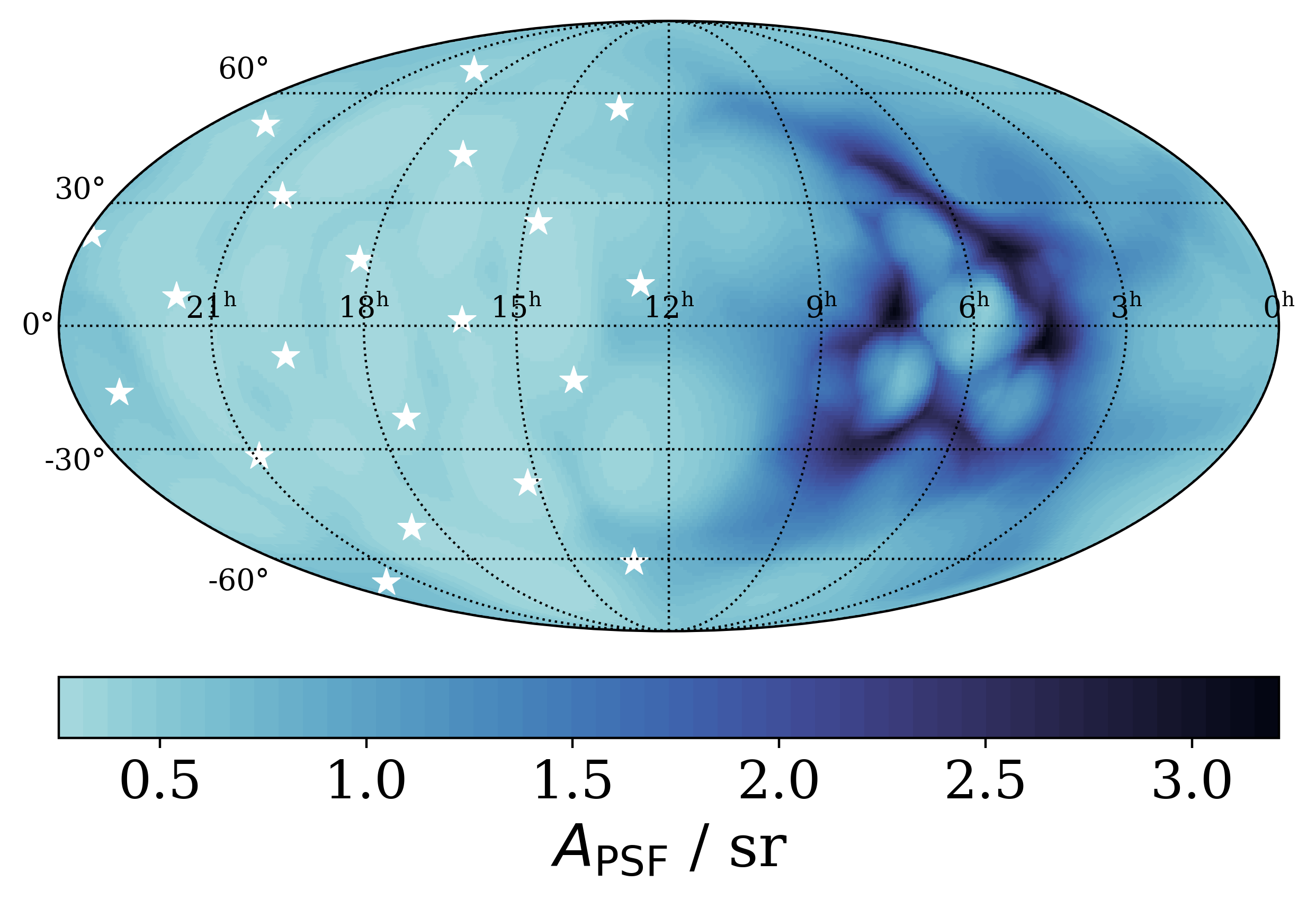}
    \includegraphics[width=0.24\textwidth]{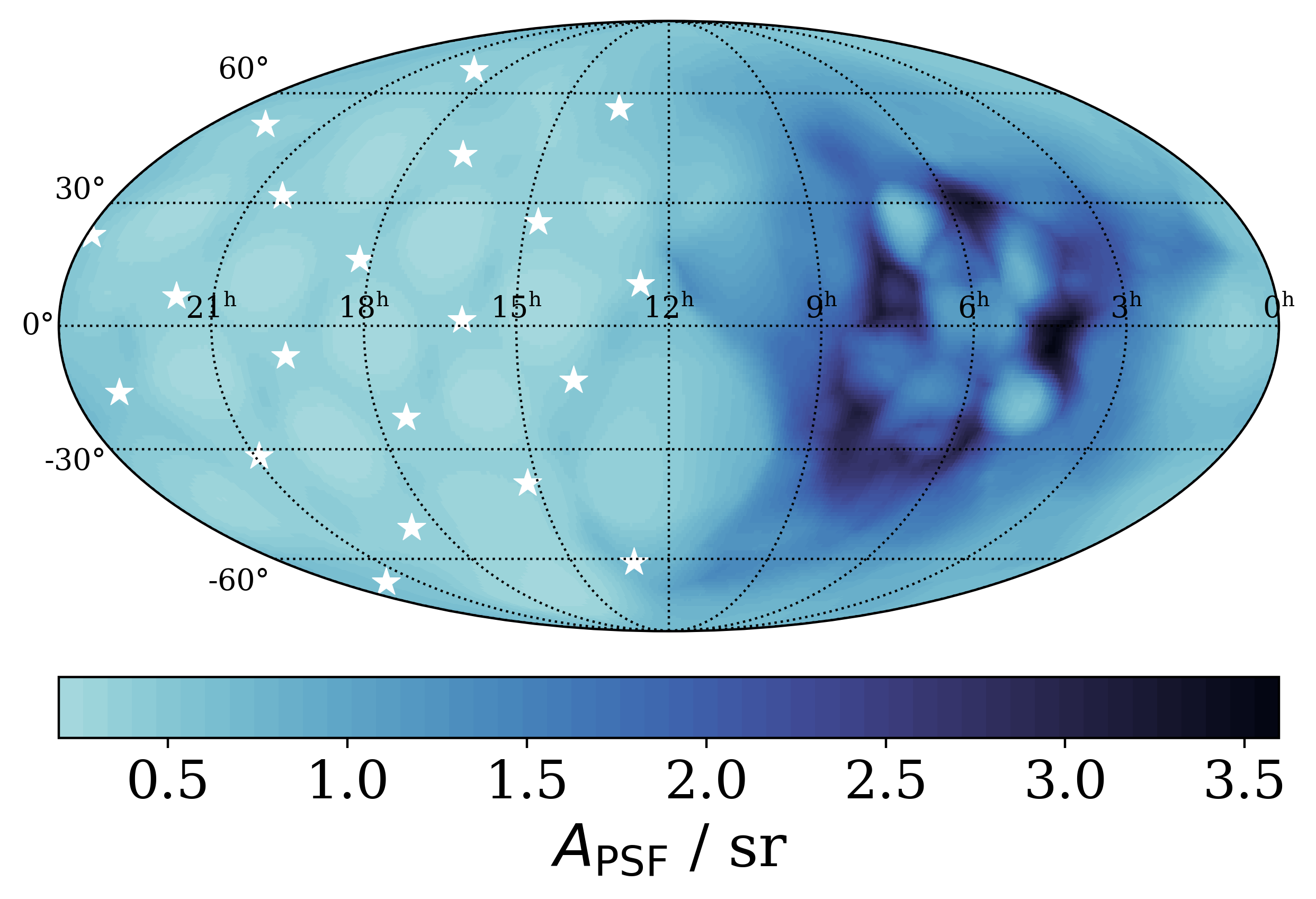}
    \includegraphics[width=0.24\textwidth]{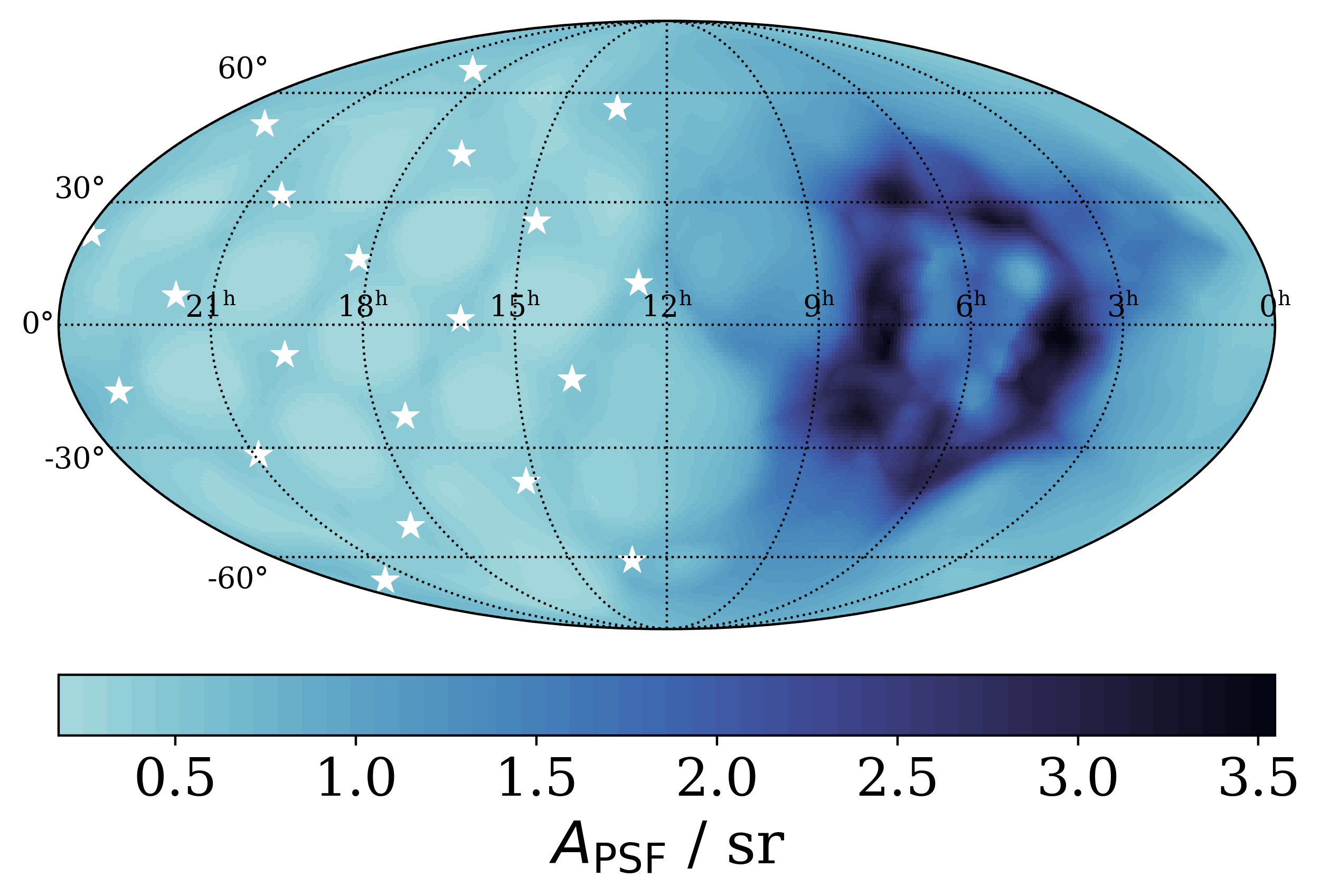}
    \caption{Point spread area sky maps for the 20-pulsar sub-array of the 40 pulsar isotropic PTA set-up. From left to right: $\lmax = \{4,8,10,12\}$. The white stars indicate the pulsar positions.}
    \label{fig:20PSR_lmax-variation}
\end{figure*}

Next we performed the same study on the 20 pulsar sub-array for $\lmax\in[4, 12]$. As we had 20 identical pulsars contributing to the array, we set $\sreg=20$. In Fig.~\ref{fig:20PSR_lmax-variation} we show exemplary sky maps for $\lmax=\{4,8,10,12\}$.

Similar to the full-array case, we find that the minimum PSF decreases with increasing $\lmax$, and that it reaches some lower bound towards $\lmax = 12$. We can also see that for $\lmax = \{8,10,12\}$ the same net-like pattern emerges as in the PSF sky maps of the full 40-pulsar PTA. More specifically, the pattern is visible on the left part of the sub-array sky maps, where the remaining 20 pulsars are located. This supports our expectation that the point source resolution of a PTA is dominated by the local pulsar constellation. Since on the left part of the map the full array and the sub-array locally follow the same constellation, the similar sky pattern demonstrates that our method suitably traces the PTA properties.

As a side note, we find that on the $\lmax = 4 $ sky map the PSF is seemingly smaller in the region with no pulsars than in the region densely populated with pulsars. Upon closer inspection, it turns out that this unintuitive result is an artefact of the area evaluation scheme, as is pointed out in Sec.~\ref{ssec:PSF_scheme}. Due to the sparse pulsar population, the corresponding clean maps only show fluctuations, but no identifiable hotspot, leading to non-meaningful estimates of the PSF in that area. This behaviour disappears when we change to a larger $\lmax$ as we use a better model for the PTA sky resolution.

\subsubsection{Minimum point spread function convergence}

We expect from the diffraction limit that any PTA cannot resolve point sources beyond the angular separation between the nearest pulsars to the source \citep{Boyle_Pen_2012}. The convergence behaviour of the PSF minimum, $A_\mathrm{PSF,min}$, that we identified in both examples above is similarly indicative of a natural resolution limit of the PTA dataset. In order to confirm the relation between the pulsar separation and the resolution limit, we compared $A_\mathrm{PSF,min}$ for different analysis set-ups (combinations of $\lmax$ and $\sreg$) to the distribution of nearest-neighbour angular separations, $\delta_\mathrm{nn}$, of the PTA. This comparison was done both for the full array (40 pulsars) and the sub-array (20 pulsars).

As the PTA PSF algorithm returns area estimates, we calculated an area-equivalent of $\delta_\mathrm{nn}$ via $A_\mathrm{nn} = 2\pi [1-\cos(\delta_\mathrm{nn}/2)]$, and used this quantity to perform the comparison. This area equivalent approximates the area between pulsars as the surface area of a spherical cap\footnote{In reality, this area is rather a polygonal structure on a spherical surface. Hence, the approximation may slightly overestimate the actual area marked out by the nearest pulsars, but it is sufficient to compare the achievable resolution to the expected diffraction limit.}, with the apex angle being $\delta_\mathrm{nn}$.

\begin{figure}
    \centering
    \includegraphics[width=\linewidth]{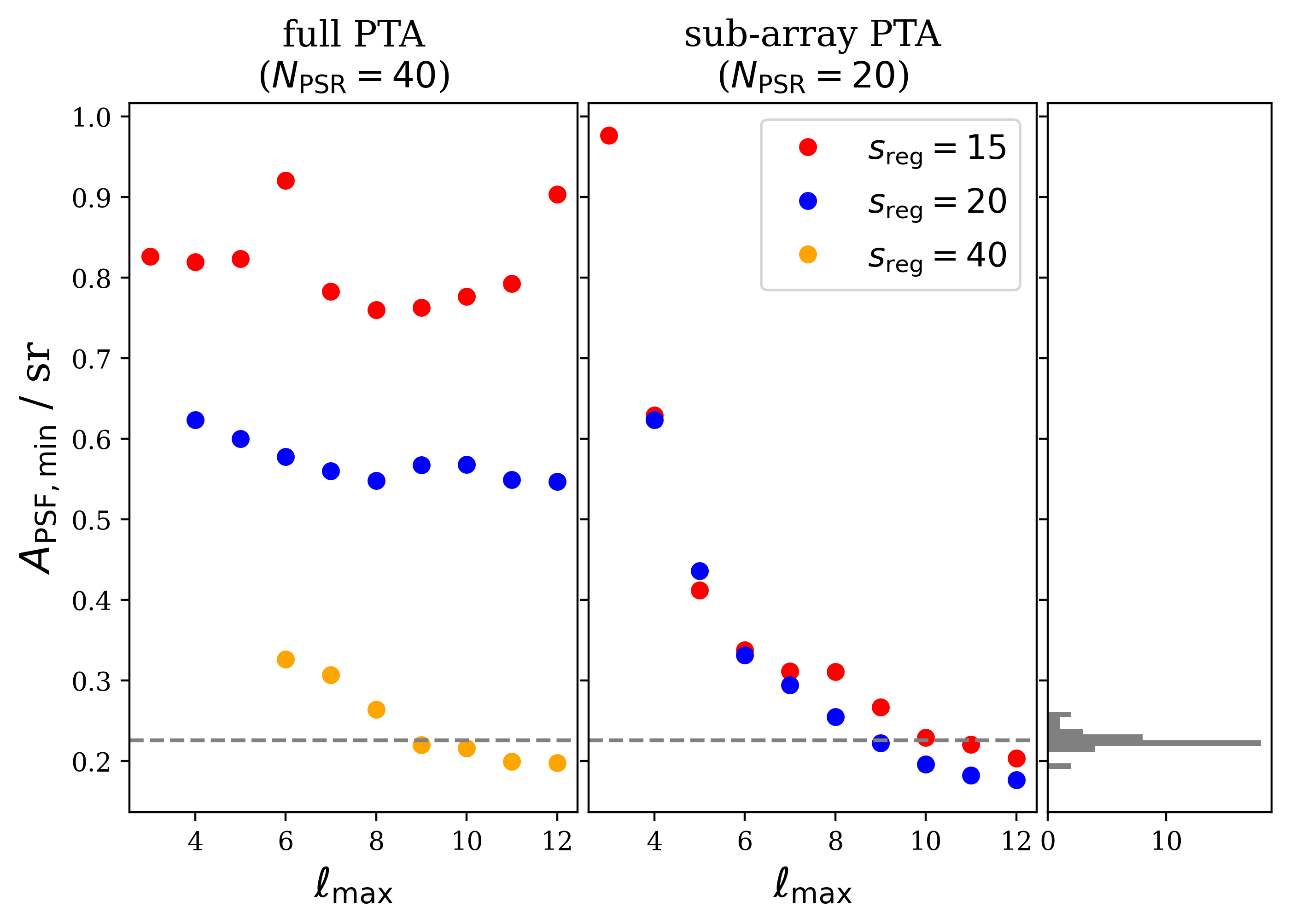}
    \caption{Minimum PSF as function of $\lmax$, for different combinations of regularisation cut-offs, $s_\mathrm{reg}$. Left: 40 PSRs. Middle: 20 PSRs. Right: Distribution of area equivalents, $A_\mathrm{nn}$, of the nearest-neighbour distances in the 40-pulsar constellation.
    }
    \label{fig:nearest-neighbour-study}
\end{figure}

The minimum PSFs of both constellations (40 and 20 pulsars) for $\lmax \in [3,11]$ and different regularisation cut-offs ($s_\mathrm{reg}=\{15,20, (40)\}$ -- red, blue, and yellow dots, respectively) are plotted in the left and middle subplot of Fig.~\ref{fig:nearest-neighbour-study}. The right subplot shows the distribution of $A_\mathrm{nn}$ for the full 40 pulsar PTA, since it is the same as the distribution of the sub-array in the densely populated part of the sky. The horizontal dashed line indicates the mean of this distribution, $\bar{A}_\mathrm{NN}$.  
For the full array, we find that varying $\lmax$ while keeping $s_\mathrm{reg}$ constant does not change $A_\mathrm{PSF,min}$. Due to the regularity of the pulsar distribution, there is minimal anisotropy across the map, and hence it is unsurprising that for $s_\mathrm{reg}=40$ the minimum PSF area resembles the angular separation distribution.
For more realistic PTAs with anisotropic pulsar constellations, both $\lmax$ and $s_\mathrm{reg}$ play a role in the minimum PSF area. 
We demonstrate by varying $\lmax$ that for a suitable value of $\sreg$, $A_\mathrm{min}$ converges towards the mean of the angular separation distribution.

\subsubsection{Determining maximum spherical harmonics degree}
\label{sssec:det_lmax}
The proposed map calculation scheme gives up on the rule of thumb to determine $\lmax$, and technically allows one to choose $\lmax$ deliberately large, as long as we properly regularise the result. In practice this is not recommended, as the dimensions of the matrices involved in the calculation depend on $N_\mathrm{modes}$, and algebraic operations may become computationally unfeasible. Hence, we need to develop a new approach to determine $\lmax$.

We established in the previous section that the maximum local resolution of a PTA is closely linked to the nearest-neighbour angular distance distribution in the PTA. Moreover, we show in Fig.~\ref{fig:nearest-neighbour-study} that the distribution mean of the area-equivalent, $\bar{A}_\mathrm{NN}$, is a conservative tracer of the expected minimum PSF achievable with the pulsar constellation at hand. 

Based on these findings, we suggest deriving $\lmax$ by comparing the mean nearest-neighbour distance from a real PTA dataset, $\Bar{\delta}_\mathrm{nn}$, to the characteristic extent of the spherical harmonics tessellation. This approach is rooted in the following geometric consideration: 
it is customary to visualise spherical harmonics via their nodal lines, chequering the sphere with parcels in which the spherical harmonic has a positive sign, and those where it has a negative sign. This tessellation of the sphere can be seen as a proxy of the resolution achievable with the respective spherical harmonic degree, $\ell$. For the modes $m = {-l,...,l}$ of a spherical harmonic of degree $l$, the mode with $m = \mathrm{round}(\ell/2)$ has the largest number of parcels across the sphere, and the closest distance of nodes. The diagonal extent of such a patch is given as $\arccos\left[\sin\left((\ell-m+2)/(\ell-m)\right)\right]$. 
Therefore, matching the diagonal parcel extent to $\Bar{\delta}_\mathrm{nn}$ returns a conservative estimate for $\lmax$.

In realistic PTA datasets, the number of pulsars effectively contributing the majority of the measured GWB signal is notably smaller than the total number of pulsars, due to different pulsar sensitivities \citep{Speri_2023}. To some degree, this is accounted for by using the distribution mean instead of the minimum nearest-neighbour distance. Clearly, proximity of pulsars alone is not guaranteed to produce a meaningful constraint on the smallest-scale modes, and it is reasonable to assume that those modes would be regularised regardless. Still, it is likely that the proposed approach leads to an overestimation of $\lmax$. More refined techniques to determine $\lmax$ could include identifying the effectively contributing pulsars using the methods presented by \cite{Speri_2023}, or studying the influence of noise on the PTA resolution, as is detailed in App.~\ref{app:generalised_PSF_formalism}. But as we provide a conservative estimate of $\lmax$, and the number of spherical harmonics modes are not the bottleneck of the anisotropy analysis. We refer these investigations to future studies.

\section{Test of the scheme with MPTA simulation} \label{sec:test_mpta}

We tested the performance of the adjusted spherical harmonics expansion scheme by applying them to simulations based on the MPTA 4.5-year dataset, similar to those presented in the appendix of \cite{MPTA2025_aniso}. 
The MPTA 4.5-year dataset comprises 83 pulsars located mostly in the southern hemisphere. The PTA constellation is indicated with the white stars in Figs.~\ref{fig:MPTA_CGW} and \ref{fig:MPTA_GWB}.
Their size reflects the inverse of their RMS; larger pulsars contribute more to the GW S/N. All pulsars were observed every 14-16 days, creating regular ToA series \citep{MPTA2025_data+noise}. 

Analogously to the simulations used in our previous work, \cite{MPTA2025_aniso}, we simulated a simplified version of the dataset, including only white noise. For each pulsar we obtained whitened ToAs from the publicly available MPTA 4.5-year ToAs and ephemerides using the \textsc{tempo2} \texttt{formIdeal} plug-in. 
This way, the simulated dataset has the same observation times, pulsar positions, and pulsar timing model as the real dataset. We then used the \texttt{libstempo.toasim} functions to add white noise fluctuations (\texttt{EFAC},\texttt{EQUAD}), as well as different GW signals to the whitened ToAs, creating two datasets:
\begin{enumerate}
    \item \textit{White noise + isotropic GWB.}
         We injected a Hellings-Downs-correlated common red noise signal with a power-law PSD into the simulated dataset. The amplitude and spectral index were chosen to match the values reported by \cite{MPTA2025_GWB}. 
            
    \item \textit{White noise + CGW.} We injected a single, monochromatic CGW signal into the dataset. In order to test the resolution in the area with a high pulsar desity, we located the supermassive black hole binary at RA~18h DEC~\SI{-45}{\degree}. We set $f_\gw = \SI{1e-8}{\hertz}$, so that the source falls between the first and second MPTA frequency bin. To achieve a significantly visible source, without giving up monochromaticity we set $\mathcal{M}_\mathrm{c}=\SI{1e8}{\msun}$ and $d_\mathrm{L} = \SI{1}{\mega\parsec}$.
\end{enumerate}

We analysed both simulations using the regularised spherical harmonics set-up, and obtained narrow-band pulsar pair cross-correlations using \textsc{defiant} \citep{Gersbach_2025}. For consistency with the MPTA 4.5-year analyses, we kept $s_\mathrm{reg}=32$, which was determined as the standard choice for the anisotropy analyses of that dataset \citep{MPTA2025_aniso}. In both cases, we set the common red noise parameters in the analysis noise specifications to $\log_{10}A_\GWB= -13.8$, $\gamma_\GWB=3.41$. For the reader familiar with the simulations performed in our previous paper \citep{MPTA2025_aniso}, we highlight that the value of $\log_{10}A_\GWB$ chosen for the anisotropy analysis in this work is slightly larger, leading to smaller S/N and $p$ values in the clean sky maps. This adjustment was necessary in order to keep the $S/N$ range of the CGW recovery for all set-ups in the well-tested range for this analysis method.

\subsection{Determination of $\lmax$}

\begin{figure}
    \centering
    \includegraphics[width=\linewidth]{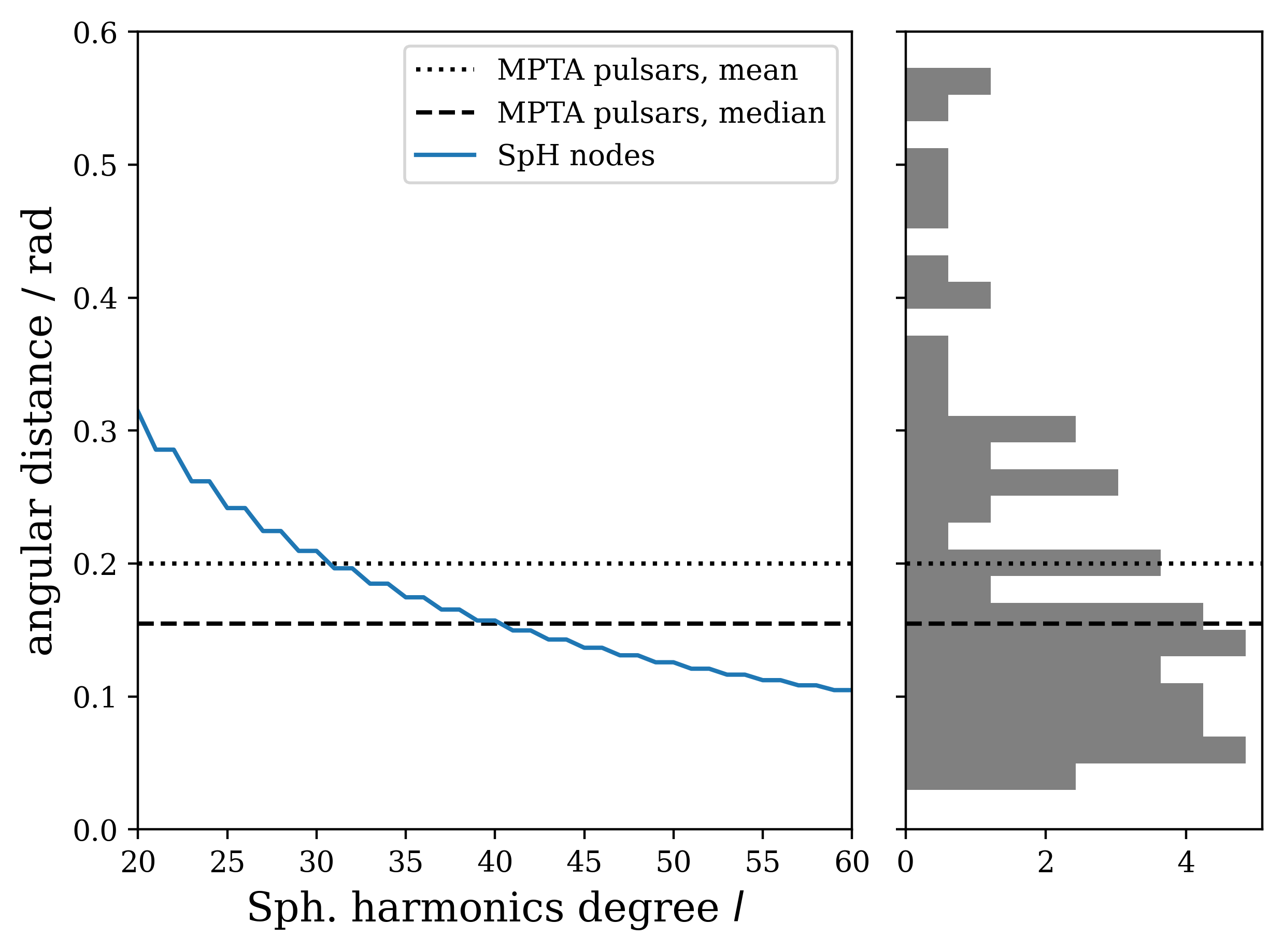}
    \caption{Estimation of the $\lmax$ based on the spherical harmonics geometry. Left: Minimum spherical harmonics tessellation extent at a given degree, $\ell$. Right: MPTA nearest-neighbour distance distribution.}
    \label{fig:MPTAresolution_sph}
\end{figure}

Using the method suggested in Section~\ref{sssec:det_lmax}, we find for the MPTA dataset $\lmax = 31$. A visualisation of the area-matching process together with the nearest-neighbour distance distribution of the MPTA is shown in Fig.~\ref{fig:MPTAresolution_sph}.
In order to shed light on the change of the result with varying $\lmax$, we decided to also analyse the datasets with $\lmax=\{16, 21\}$.

\begin{figure}
    \centering
    \includegraphics[width=\linewidth]{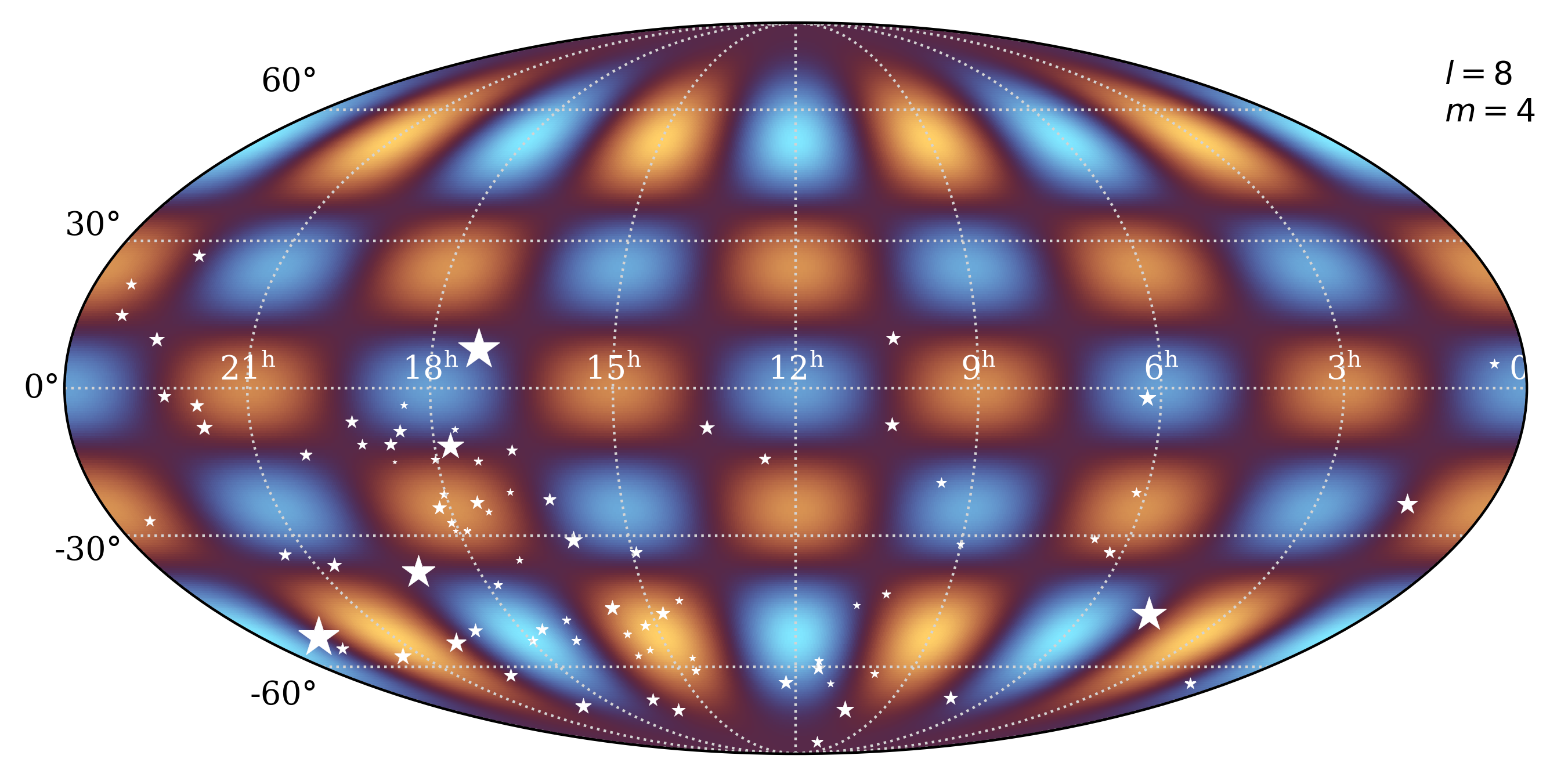}
    \includegraphics[width=\linewidth]{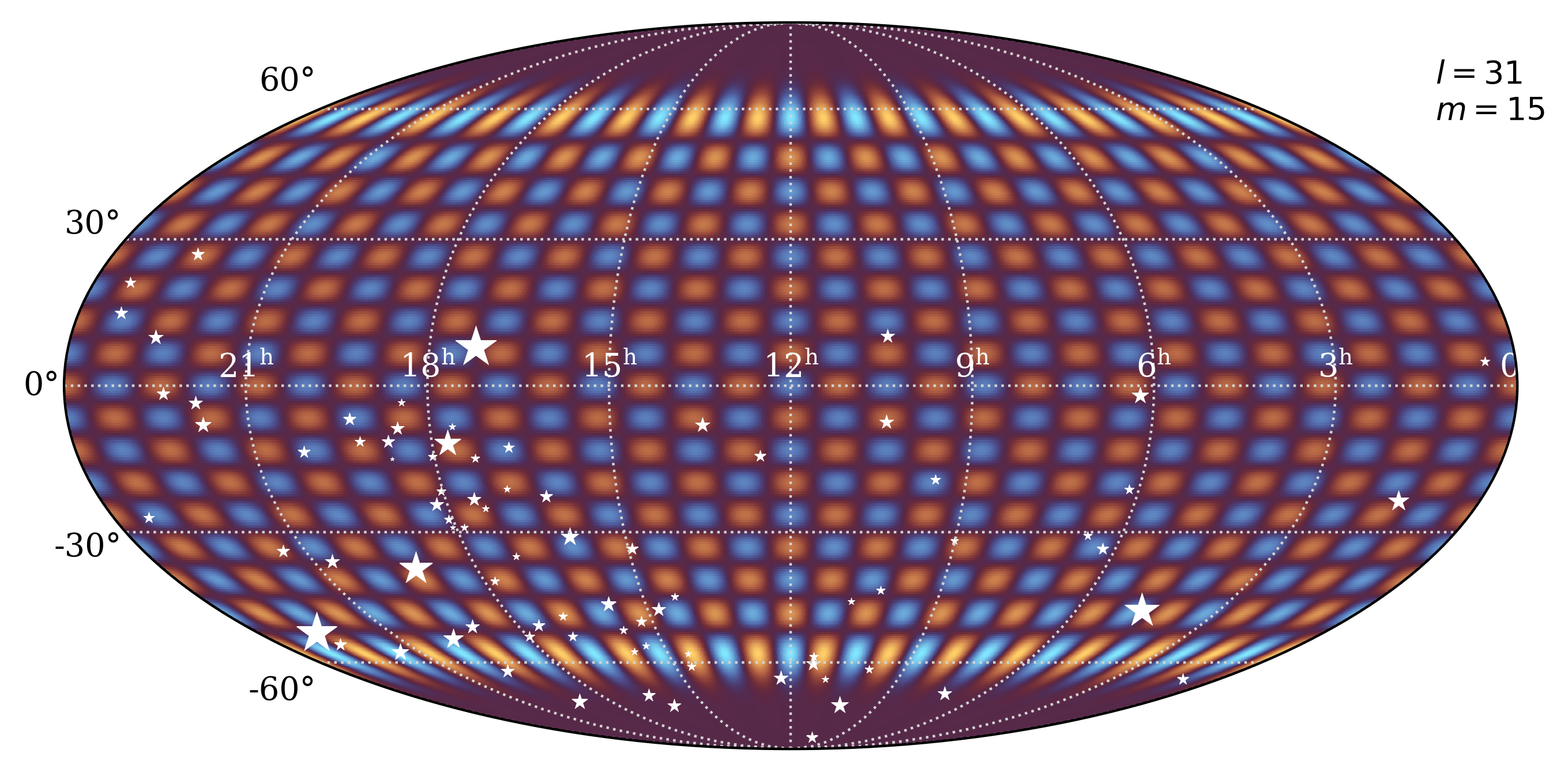}
    \caption{Visual comparison of the spherical harmonics tessellation to the MPTA pulsar constellation. Top: $\lmax=8$. Bottom: $\lmax=31$.}
    \label{fig:MPTA_with_SpH}
\end{figure}

A representation of the maximum spherical harmonics tessellation achieved with these $\lmax$, in comparison to the under-resolving tessellation of $\lmax=8$, is shown in Fig.~\ref{fig:MPTA_with_SpH}. The sky maps show the spherical harmonic mode with the largest number of knots, as well as the position of the MPTA pulsars as white stars.

\subsection{Results}

\begin{figure*}
    \centering
    \includegraphics[width=0.245\linewidth]{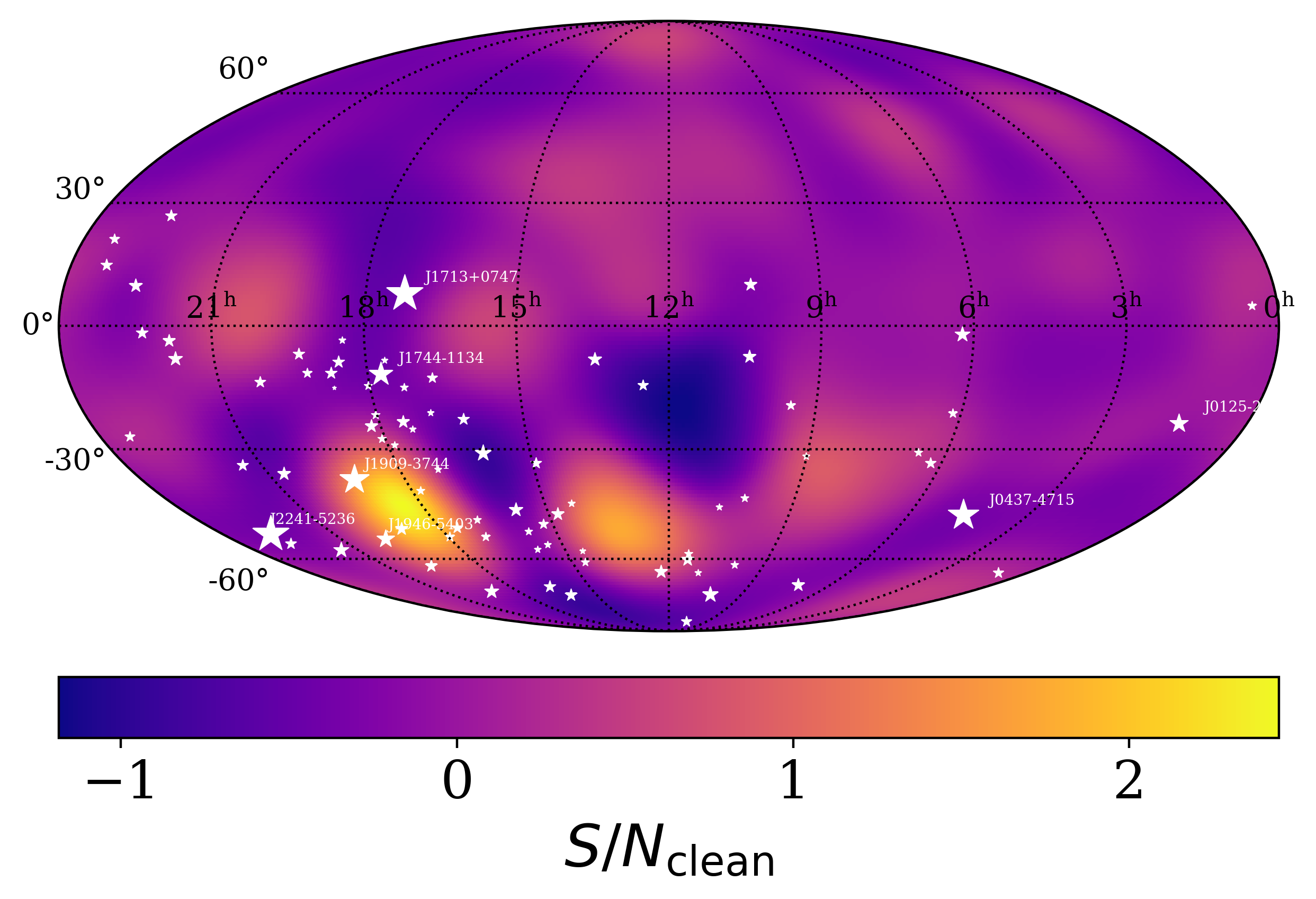}
    \includegraphics[width=0.245\linewidth]{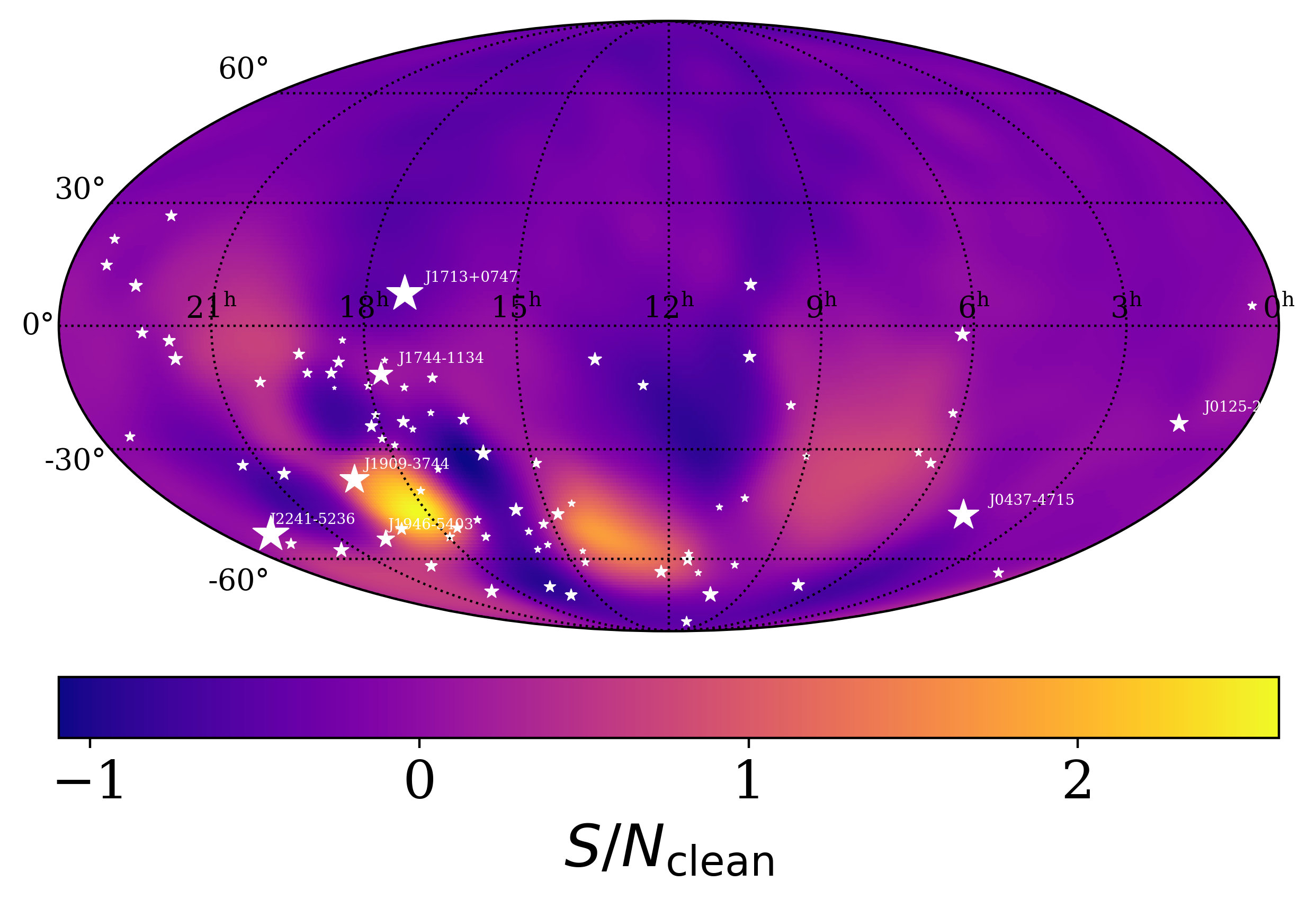}
    \includegraphics[width=0.245\linewidth]{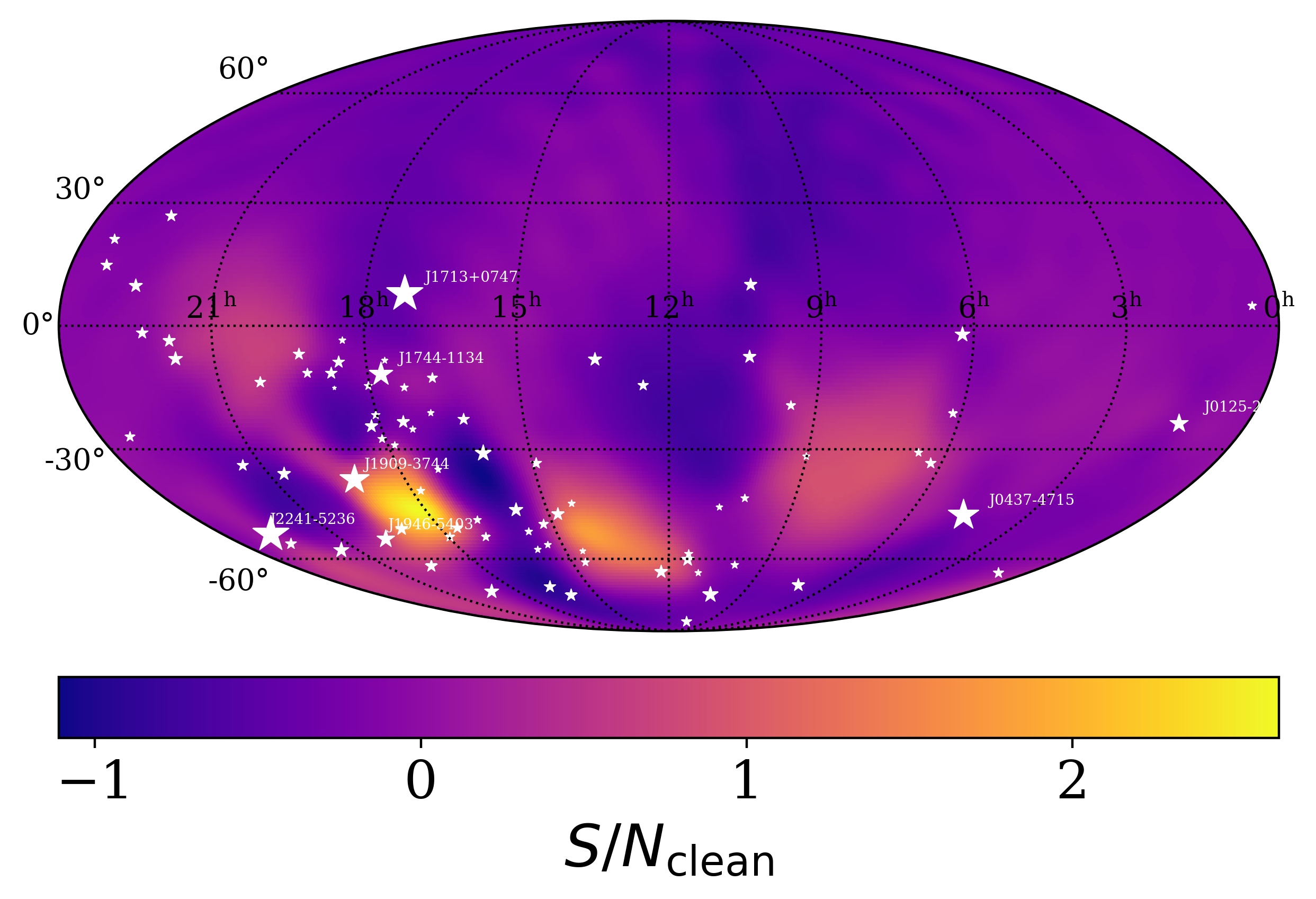}
    \includegraphics[width=0.245\linewidth]{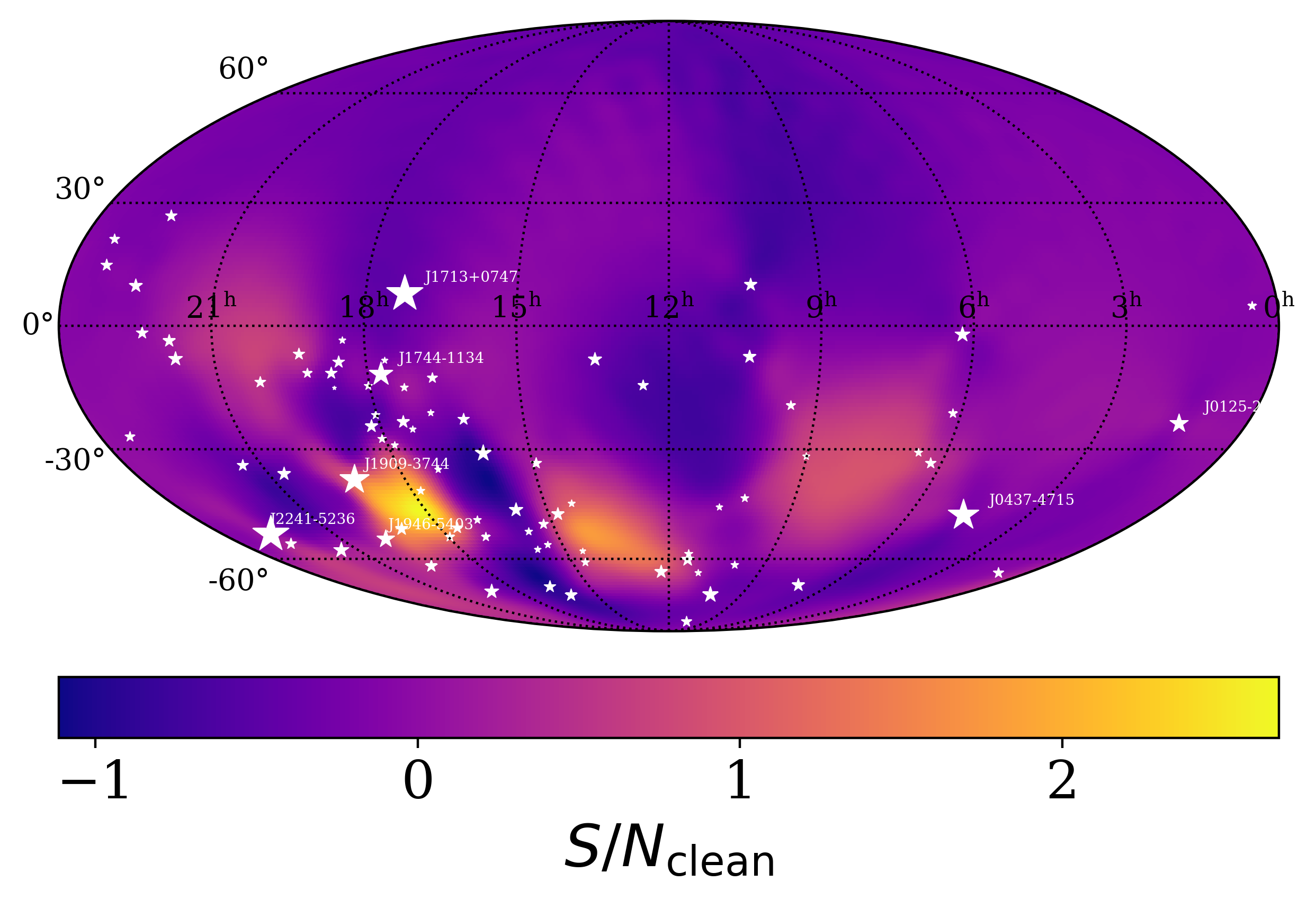}
    \caption{Regularised, clean S/N maps of the first frequency bin (\SI{7}{\nano\hertz}) for the simulated MPTA datasets containing WN and a single CGW source at RA~18h DEC~\SI{-45}{\degree}. From left to right, the spherical harmonics expansion is set to $\lmax = \{8,16,21,31\}$. For all maps, the regularisation cut-off is set to $s_\mathrm{reg} = 32$.}
    \label{fig:MPTA_CGW}
\end{figure*}

We analysed both simulations with $\lmax=\{8,16,21,31\}$ and set $s_\mathrm{reg}=32$ to maintain comparability to the simulation analyses in our previous work \citep{MPTA2025_aniso}. 
The clean $S/N$ maps of the CGW recovery are shown in Fig.~\ref{fig:MPTA_CGW}. The visibly reduced area and blur of the hotspot demonstrate that increasing $\lmax$ beyond \num{8} leads to a better constraint on the source location. With $\lmax=8$, the recovered hotspot has an angular size of \SI{0.25}{\sr}; for $\lmax=31$, the size decreases to \SI{0.15}{\sr}. 
Simultaneously, the increased maximum of the colour bar indicates a slight increase in the maximum $S/N$ of the hotspot. This shows that the finer spherical harmonics tessellation allows us to exclude additional noise in the relevant spherical harmonics modes, which was broadening the hotspot in the coarser expansion before. In all three cases, the S/N, the corresponding $p$ value, the hotspot area, and the position of the pixel with maximum $S/N$ are collected in Tab.~\ref{tab:MPTA_rec_vals}. We also find that overall the localisation of the hotspot improves. With increasing $\lmax$, the right ascension of the recovered hottest pixel corresponds significantly better to the CGW position (changed from 18h24m to 18h04m for $\lmax=31$), but it becomes slightly offset in declination (changed from \SI{-45}{\degree} to \SI{-46}{\degree} for $\lmax=31$). We also notice that no artificial hotspots appear at higher $\lmax$, which demonstrates the stability of our approach.

Studying the map evolution between $\lmax=16$ and $31$, we find the previously explored convergence behaviour towards a minimal local resolution. The maximum $S/N$ and the spatial resolution of the hotspot do not improve significantly. We only find more pronounces traces of the residual diffraction pattern on the northern hemisphere.
Looking closer at the extent of the hotspot of the maps in Fig.~\ref{fig:MPTA_CGW}, we find that they behave in the manner described by \cite{Boyle_Pen_2012}. The resolution of a point source, or similarly the extent of the hotspot, is given by the smallest quadrilateral of pulsars surrounding the source position.

\begin{figure*}
    \centering
    \includegraphics[width=0.245\linewidth]{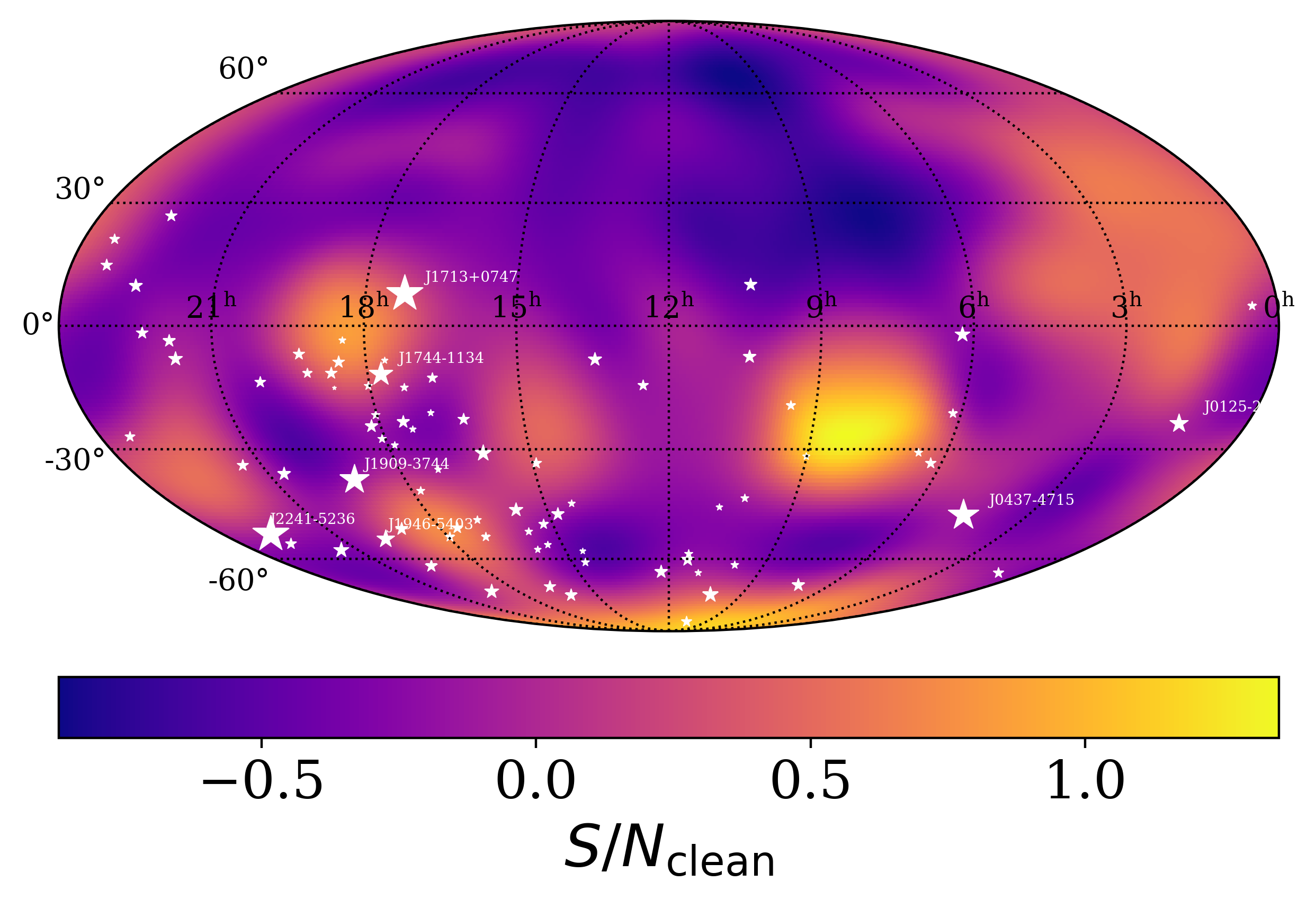}
    \includegraphics[width=0.245\linewidth]{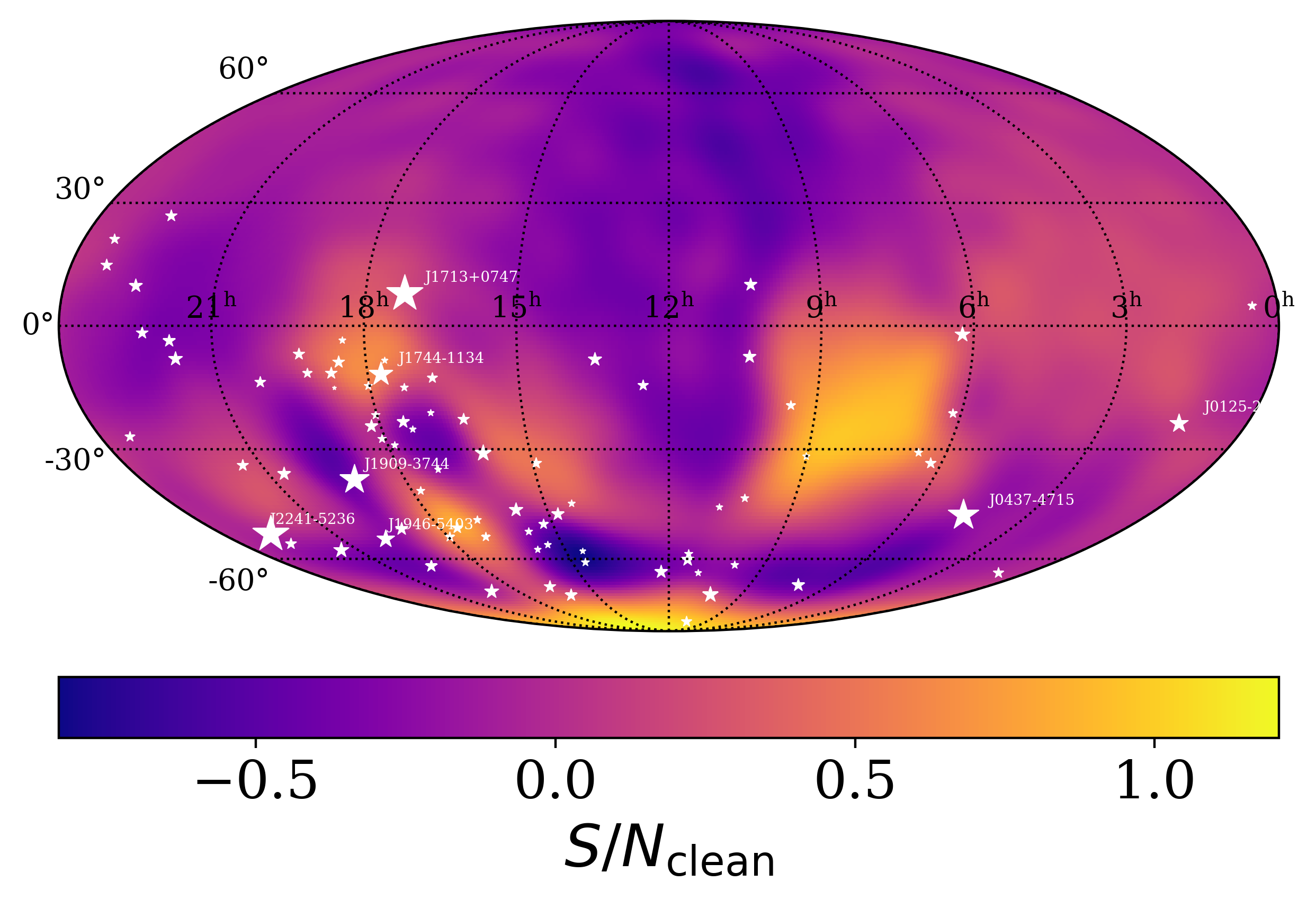}
    \includegraphics[width=0.245\linewidth]{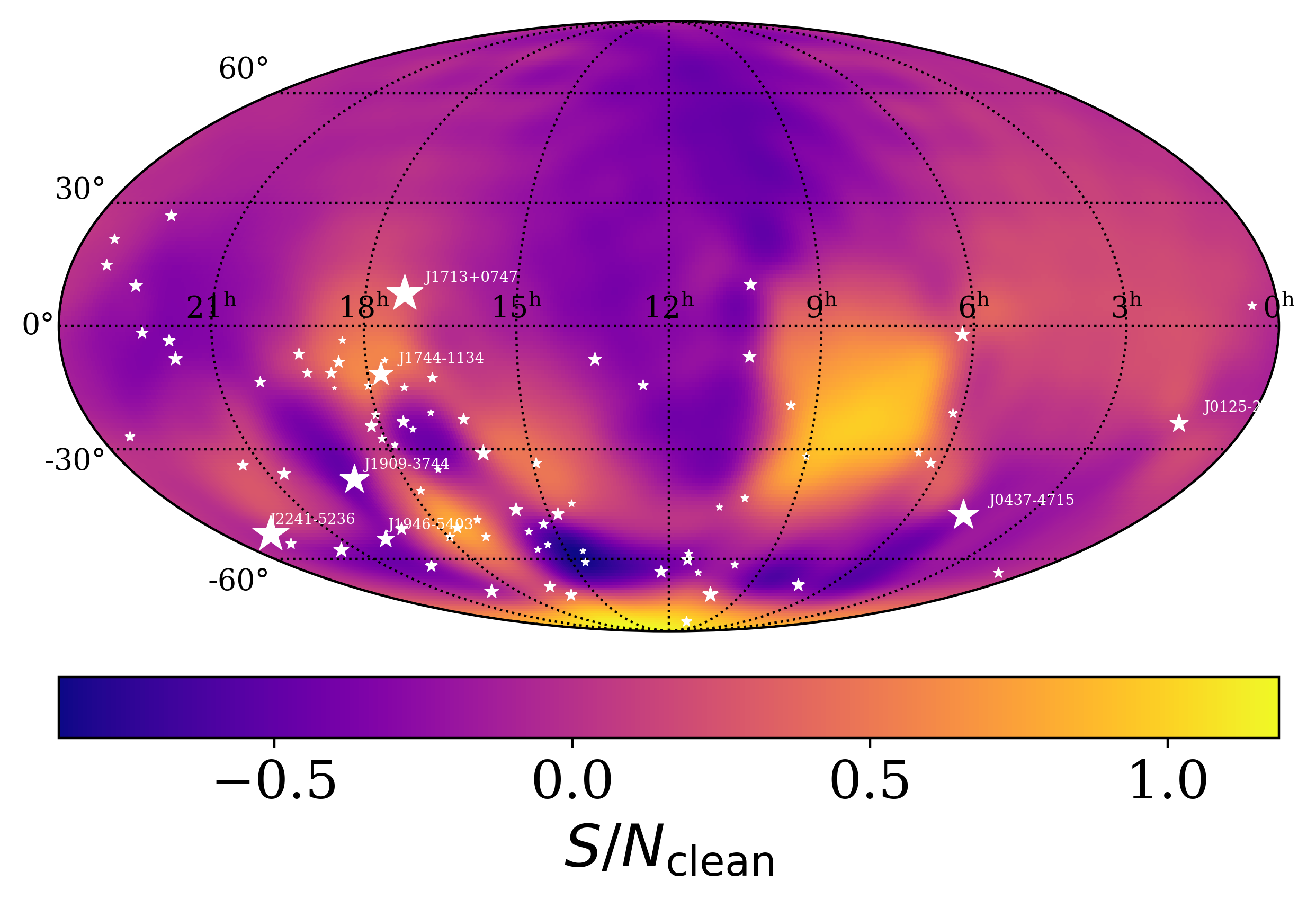}
    \includegraphics[width=0.245\linewidth]{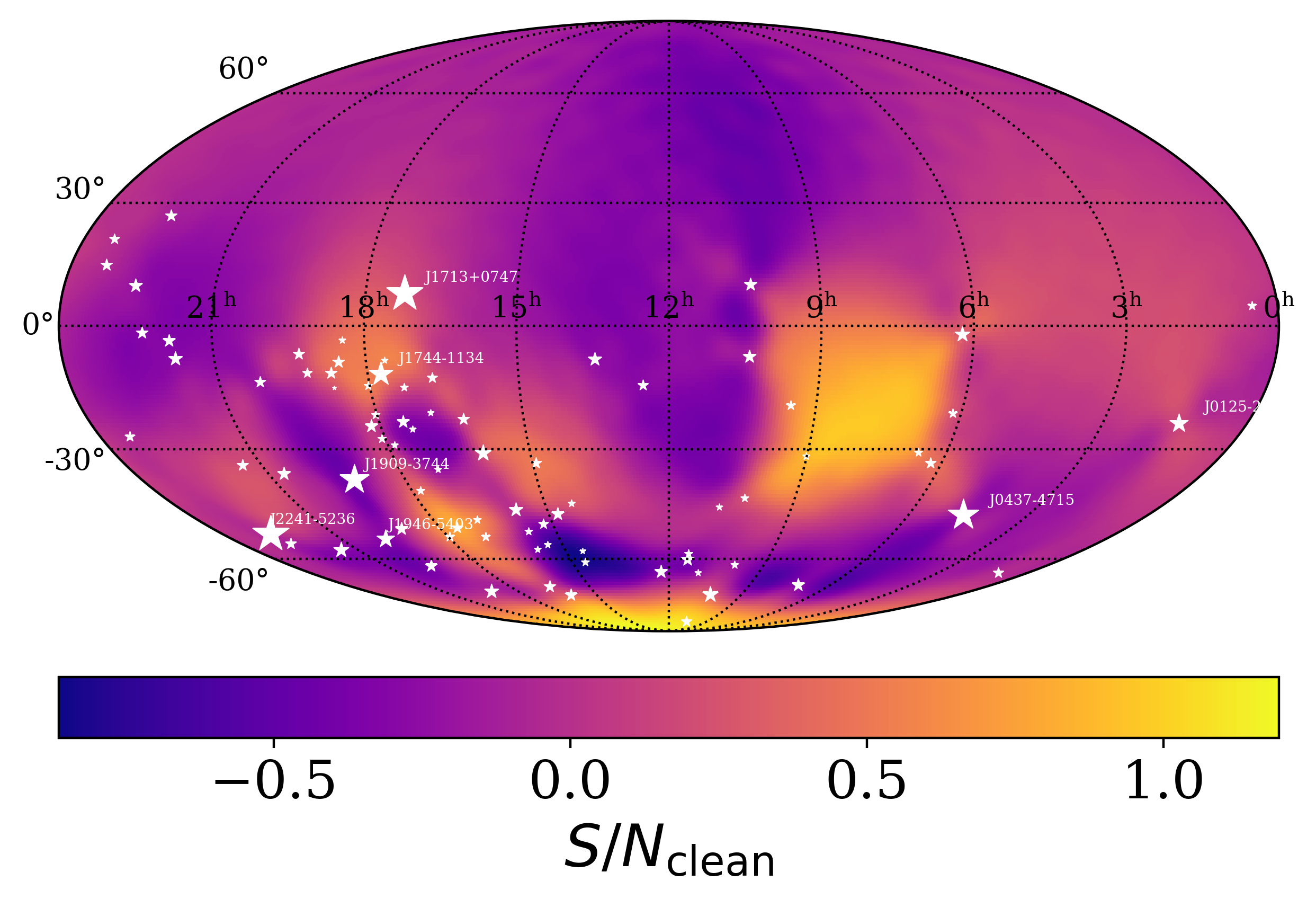}
    \caption{Regularised, clean S/N maps of the first frequency bin (\SI{7}{\nano\hertz}) for the simulated MPTA datasets with WN and isotropic GWB injection. From left to right, the regularised spherical harmonics analysis is set up with to $\lmax = \{8,16,21,31\}$ and $s_\mathrm{reg} = 32$.}
    \label{fig:MPTA_GWB}
\end{figure*}

Similarly, we analysed a simulation containing an isotropic GWB injection. The resulting clean $S/N$ maps, displayed in Fig.~\ref{fig:MPTA_GWB}, illustrate that varying the spherical harmonics expansion does not have a significant influence on the maps' maximum $S/N$. Additionally, it can be seen by eye that a larger $\lmax$ makes the recovery look more isotropic in sky areas with a sparse pulsar population. The maximum $S/N$ values and corresponding $p$ values are collected in Tab.~\ref{tab:MPTA_rec_vals} as well. Interestingly, $S/N_\mathrm{max}$ even decreases with larger $\lmax$. This is another manifestation of how the more realistic local resolution mitigates signal confusion due to noise being collected across overly large patches. Across all three spherical harmonics expansions, the $p$ values are similar, consistent with an isotropic background. Most importantly, this means that our method does not give rise to artificially large $S/N$ values in the absence of any actual anisotropy.

\begin{table}
    \caption{Characteristic quantities of the MPTA simulations' anisotropy analyses.}
    \centering
    \begin{tabular}{c||c|c|c||c|c}
            & \multicolumn{3}{l||}{CGW} & \multicolumn{2}{l}{GWB}  \\ \hline
    $\lmax$ & $S/N$ &$p$-val. & $A$ /sr & $S/N$ & $p$-val.          \\ \hline\hline
        8   & \num{2.45} & \num{0.186} & 0.25 & \num{1.35} & 0.912\\
        16  & \num{2.61} & \num{0.011} & 0.17 & \num{1.21} & 0.956\\
        21  & \num{2.63} & \num{0.003} & 0.16 & \num{1.19} & 0.971\\
        31  & \num{2.67} & \num{0.001} & 0.15 & \num{1.19} & 0.978
    \end{tabular}
    \tablefoot{ The first four columns contain the maximum clean map $S/N$, its $p$ value, and the position and the hotspot area for the CGW injection at RA~18h DEC~\SI{-45}{\degree}. The last two columns collect the maximum $S/N$ and $p$ values for the GWB injection.}
    \label{tab:MPTA_rec_vals}
\end{table}

\section{Summary and outlook} \label{sec:summary}

This work builds on our previous publication \citep{MPTA2025_aniso} and contains two major results.
We develop a framework to evaluate the PSF of a PTA dataset depending on the choice of parameters of the anisotropy analysis. 
As a second result, we demonstrate with this framework that the regularisation scheme allows us to alter the spherical harmonics expansion of the anisotropy analysis, such that it reflects the local PTA resolution more realistically. We hence present a reworked anisotropy analysis scheme, which can be summarised in two steps:
 \begin{enumerate}
    \item We determine suitable values for $\lmax$ (the maximum spherical harmonic degree of the expansion) based on the mean or median of the distribution of pulsars in the PTA.
    \item We remove poorly unconstrained modes by imposing a regularisation threshold, so that the number of modes is comparable to (or smaller than) the number of pulsars: \ $\sreg \leq N_\mathrm{PSR}$. For real PTA datasets, the choice of $\sreg$ is also influenced by the quality of the pulsars \citep{MPTA2025_aniso}, such that it is likely smaller than $N_\mathrm{PSR}$ (for example, $\sreg$ was chosen to be 32 for the MPTA 4.5-year analysis, due to the noise and sensitivity properties of the dataset).
\end{enumerate}

We tested this scheme with simulations based on the MPTA 4.5-year PTA dataset. Analysing a continuous-wave injection, we achieved a factor of two improvement in the statistical significance compared to the method from \cite{MPTA2025_aniso}. On top of this, we improved the area constraint of the corresponding hotspot by 40\% compared to the classical scheme. This demonstrates that in areas with high pulsar densities the better reflected PTA resolution leads to a significantly refined localisability of hotspots.

We recommend that future pulsar timing anisotropy analyses use this updated scheme to set up the analysis parameters. Aside from the improved statistical significance of a potential anisotropy, optimised PTA sky maps connect naturally to targeted searches for individual supermassive black hole binaries, as they facilitate follow-up studies.

The current bottleneck of the analysis is the calculation of the pulsar-pair cross correlations, so as PTAs grow, an efficient implementation of this step is key to a fast anisotropy analysis. Forthcoming studies can also focus on fine-tuning the scheme for estimating $\lmax$, as is detailed in Sec.~\ref{sssec:det_lmax}.

Furthermore, there are a number of applications of the PSF framework beyond the scope of this paper. Future work may target the impact of noise misspecification on angular resolution, the interplay between residual noise and the regularisation scheme, and the sky position recovery of point sources. It may also be interesting to investigate whether it is possible to optimise PTA set-ups such that the corresponding sky maps target regions of high interest; for example, the Virgo cluster or around continuous-wave candidates.

\section*{Data availability}
The code developed and used for this work can be accessed under \url{https://github.com/mpta-gw/cartography.git}.

\begin{acknowledgements}
KG, DC and MK acknowledge continuing valuable support from the Max-Planck society. KG acknowledges support from the International Max Planck Research School (IMPRS) for Astronomy and Astrophysics at the Universities of Bonn and Cologne.
ET and RSN acknowledge support from the Australian Research Council (ARC) Centres of Excellence CE170100004 and CE230100016 and ARC DP230103088.
MK acknowlegdes support by the CAS-MPG legacy Programme.
MM acknowledges support from the NANOGrav Collaboration's National Science Foundation (NSF) Physics Frontiers Center award numbers 1430284 and 2020265.

This publication made use of open source python libraries including \textsc{numpy} \citep{numpy}, \textsc{scipy} \citep{scipy}, \textsc{matplotlib} \citep{matplotlib}, and \textsc{healpy} \citep{Gorski_2005}, as well as the python-based pulsar analysis packages, \textsc{libstempo}, \textsc{enterprise} \citep{enterprise}, \textsc{enterprise\_extensions} \citep{enterprise_extensions} and \textsc{defiant} \citep{Gersbach_2025}.
\end{acknowledgements}

\bibliographystyle{aa}
\bibliography{refs}

\appendix
\section{Extension of the point spread formalism to evaluate noise misspecification}
\label{app:generalised_PSF_formalism}

Despite growing efforts to model the noise budget of the pulsars in a PTA dataset as accurately as possible, all PTA data analysis is prone to noise misspecification, i.e.\ an unmodelled noise budget left in the data. For example, comparing the noise models used in the EPTA DR2 GWB analysis \citep{EPTA_DR2_GWB} to those employed in the MPTA 4.5-year GW analysis \citep{MPTA2025_data+noise}, we find an increasing variety of noise processes found in datasets, mainly owing to an improved ToA precision and higher observing frequency bandwidths. This raises the question, if these noise contributions have been present in the EPTA dataset yet remained unnoticed. After all, there is currently no way to absolutely determine the precise and accurate noise model of every pulsar. 

It is safe to assume that residual noise also affects the sensitivity of a PTA to potential GWB anisotropies. We can expect it to contribute to the blurring of a single source during the recovery, increasing the size of the PSF. Hence it is desirable to ultimately include this effect in the PSF calculation framework developed in Sec.~\ref{sec:sky_resolution}. 

In the frequentist anisotropy analysis, the noise budget enters the clean map calculation via the Fisher matrix $\mtx{M}$, as it contains the cross-correlation covariance matrix $\mtx{\Sigma}$, whose entries are determined from the noise parameters (e.g. the white noise parameters \textsc{EFAC}, \textsc{EQUAD}, the red noise spectral parameters $A$ and $\gamma$, etc.). 

We can generalise the PTA PSF calculation scheme to allow for a hypothetical noise misspecification by altering the Fisher matrices entering Eq.~\eqref{eq:hot_pixel_recovery}. The Fisher matrix used to construct the dirty map of the hot pixel, now denoted by $\mtx{\mathcal{M}}$, contains the `true' noise values in its true cross-correlation covariance matrix, $\mtx{\mathcal{S}}$. The (regularised) inverted Fisher matrix, $\Tilde{\mtx{M}}^{-1}$, is calculated with different noise values, representing the noise misspecification. The clean map recovered for a single hot pixel is now given as:
\begin{equation}
    \Vec{\mathcal{P}}_i' = \mtx{U}\Tilde{\mtx{M}}^{-1} \Vec{X}_i = \mtx{U}\Tilde{\mtx{M}}^{-1} \mtx{\mathcal{M}} \mtx{U}^\dagger \Vec{P}_{\hat{\Omega},i}.\label{eqapp:realistic_clean_pix}
\end{equation}

While Eq.~\eqref{eqapp:realistic_clean_pix} now fully depicts a realistic recovery process, it is desirable to single out the individual contributions of both the regularisation procedure and a potential noise misspecification to the broadening of the hot pixel in the recovered clean map, $\Vec{P}_{\hat{\Omega},i}'$. To this end we rearrange $\Tilde{\mtx{M}}^{-1} \mtx{\mathcal{M}}$ in Eq.~\eqref{eqapp:realistic_clean_pix}, as
\begin{align}
    \Tilde{\mtx{M}}^{-1} \mtx{\mathcal{M}} &= \Tilde{\mtx{M}}^{-1} \; \mtx{M}\: \mtx{M}^{-1} \;\mtx{\mathcal{M}} \\
    & = \mtx{\Lambda}_{\sreg}\cdot \mtx{M}^{-1}\mtx{\mathcal{M}} \\
    & = \mtx{\Lambda}_{\sreg}\cdot\mtx{\mathfrak{N}},\label{eqapp:noise_and_reg_contr}
\end{align}
factorising it into the regularisation contribution (cf.\ Eq.\eqref{eq:lambda_def}) and the noise misspecification contribution, which we defined as $\mtx{\mathfrak{N}} \equiv \mtx{M}^{-1}\mtx{\mathcal{M}}$. 

Using the definition of the Fisher matrix, $\mtx{M} \equiv \mtx{R}^\dagger\mtx{\Sigma}^{-1}\mtx{R}$, we find an alternative expression for the noise misspecification contribution,
\begin{align}
    \mtx{\mathfrak{N}} = \mtx{M}^{-1}\mtx{\mathcal{M}} &= \mtx{R}^{-1} \mtx{\Sigma} {\mtx{R}^\dagger}^{-1} \;\;\mtx{R}^\dagger \mtx{\mathcal{S}}^{-1}\mtx{R} \\
    &= \mtx{R}^{-1}\mtx{\Sigma}\mtx{\mathcal{S}}^{-1}\mtx{R},
\end{align}
emphasising the role of the pulsar-pair cross covariance matrix in the noise model contribution to the PTA PSF.

Putting together Eqns.~\eqref{eqapp:realistic_clean_pix} and \eqref{eqapp:noise_and_reg_contr}, the general expression for the PSF calculation now reads:
\begin{equation}\label{eqapp:hot_pixel_noisemisspec_factorised}
    \Vec{P}_{\hat{\Omega},i}' = \mtx{U} \; \mtx{\Lambda}_s\mtx{\mathfrak{N}} \; \mtx{U}^\dagger \; \Vec{P}_{\hat{\Omega},i}.
\end{equation}
This factorisation demonstrates a couple of noteworthy points:
Even without applying a regularisation scheme ($\mtx{\Lambda}_{\sreg} = \mathds{1}$), a point source is blurred due to residual noise in the dataset. More broadly speaking, any deviation of both $\mtx{\Lambda}_s$ and $\mtx{\mathfrak{N}}$ from a unit matrix gives rise to a PSF. Furthermore we observe that we can consistently recover Eq.~\eqref{eqapp:realistic_clean_pix} assuming perfect noise models, i.e. $\mtx{\mathfrak{N}} = \mathds{1}$, as indicated in Sec.~\ref{ssec:distortion_matrix}. Finally we note that --- being read from right to left according to the matrix multiplication --- Eq.~\eqref{eqapp:hot_pixel_noisemisspec_factorised} is satisfyingly a mathematical reflection the loss of information throughout the anisotropy analysis: First due to noise misspecification, then due to the regularisation.

\end{document}